\documentclass[12pt]{article}
\usepackage[OT1]{fontenc}
\usepackage[utf8]{inputenc}
\usepackage{tocloft}
\usepackage{amsfonts}
\usepackage{amsmath}
\usepackage{amssymb}
\usepackage{array}
\usepackage{slashed}
\usepackage{bigints}
\usepackage{pifont}
\usepackage{bm}
\usepackage{bbm}
\usepackage{upgreek}
\usepackage{booktabs}
\usepackage[nosort]{cite}
\usepackage[table,dvipsnames]{xcolor}
\usepackage{dsfont}
\usepackage{float}
\usepackage{framed}
\usepackage{graphicx}
\usepackage{indentfirst}
\usepackage{mathrsfs}
\usepackage{multirow}
\usepackage{pdflscape}
\usepackage{setspace}
\usepackage{subdepth}
\usepackage{subfig}
\usepackage{titlesec}
\usepackage{wrapfig}
\usepackage[all]{xy}
\usepackage{young}
\usepackage[vcentermath]{youngtab}
\usepackage{relsize}
\usepackage{stackengine}
\usepackage{verbatim}
\usepackage{slashed}
\usepackage{cancel}
\usepackage{tikz}
\usetikzlibrary{arrows,decorations.markings,cd}

\newcommand{\rd}{{\rm d}}

\usepackage{hyperref}
\hypersetup{colorlinks=true}
\hypersetup{linkcolor=black}
\hypersetup{citecolor=black}
\hypersetup{urlcolor=black}

\numberwithin{equation}{section}

\usepackage[left=2.5cm,right=2.5cm,top=2.5cm,bottom=3cm]{geometry}
\fontdimen2\font=1.2\fontdimen2{\jot}{5pt}

\newlength{\bibitemsep}
\newlength{\bibparskip}
\let\oldthebibliography\thebibliography
\renewcommand\thebibliography[1]{%
  \oldthebibliography{#1}%
  \setlength{\parskip}{\bibitemsep}%
  \setlength{\itemsep}{\bibparskip}%
}

\titleformat{\section}{\bfseries}{\thesection.}{4pt}{}
\titlespacing{\section}{0pt}{20pt}{6pt}

\titleformat{\subsection}{\normalfont\itshape}{\thesubsection.}{4pt}{}
\titlespacing{\subsection}{0pt}{15pt}{6pt}

\titleformat{\subsubsection}{\normalfont\itshape}{\thesubsubsection.}{4pt}{}
\titlespacing{\subsubsection}{0pt}{15pt}{6pt}

\titleformat{\paragraph}{\normalfont\itshape}{\theparagraph.}{4pt}{}
\titlespacing{\paragraph}{0pt}{15pt}{6pt}

\renewcommand{\tilde}{\widetilde}

\renewcommand{\hat}{\widehat}

\DeclareMathOperator{\tr}{tr}
\DeclareMathOperator{\Tr}{Tr}

\DeclareMathAlphabet{\mathbfsf}{OT1}{cmss}{bx}{n}

\newcommand{\N}{{\mathbb N}}
\newcommand{\Z}{{\mathbb Z}}

\newcommand{\R}{{\mathbb R}}

\newcommand{\mb}[1]{\mathbf{#1}}

\newcommand{\ms}[1]{\mathscr{#1}}
\newcommand{\mf}[1]{\mathfrak{#1}}

\newcommand{\SL}{\mathscr{L}}

\newcommand{\bR}{\mathbb{R}}
\newcommand{\bZ}{\mathbb{Z}}

\newcommand{\cA}{\mathcal{A}}
\newcommand{\cB}{\mathcal{B}}

\newcommand{\cE}{\mathcal{E}}
\newcommand{\cF}{\mathcal{F}}

\newcommand{\cI}{\mathcal{I}}

\newcommand{\cM}{\mathcal M}
\newcommand{\cN}{\mathcal{N}}
\newcommand{\cO}{\mathcal{O}}

\newcommand{\cP}{\mathcal{P}}
\newcommand{\cQ}{\mathcal Q}

\newcommand{\cT}{\mathcal{T}}
\newcommand{\cV}{\mathcal{V}}

\newcommand{\cY}{\mathcal{Y}}

\newcommand{\su}{\mf{su}}

\newcommand{\ed}{\,.}
\newcommand{\ec}{\,,}

\newcommand{\scrL}{\mathscr{L}}

\newcommand{\be}{\begin{equation}}
\newcommand{\ee}{\end{equation}}
\newcommand{\beq}{\begin{equation}}
\newcommand{\eeq}{\end{equation}}

\newcommand{\zero}{^{(0)}}

\newcommand{\one}{^{(1)}}

\newcommand{\gp}{^{(p)}}

\newcommand{\ii}{{\rm i}}
\newcommand{\e}{{\rm e}}

\allowdisplaybreaks 

\DeclareFontShape{OT1}{cmr}{mx}{n}%
{<->cmr10}{}
\newcommand{\mytitlefont}{\fontseries{mx}\selectfont}
\DeclareMathAlphabet{\titlemath}{OT1}{cmr}{mx}{n}

\begin{document}

%
\begin{titlepage}
\begin{center}
~\\[1.5cm]
{\fontsize{27pt}{0pt} \mytitlefont Comments on Global Symmetries and Anomalies of $5d$ SCFTs}
~\\[1.25cm]
Pietro Benetti Genolini\,$^{1,2}$ and Luigi Tizzano\,$^{3,4}$\hskip1pt
~\\[0.5cm]
{$^1$~{\it Department of Applied Mathematics and Theoretical Physics,\\
University of Cambridge, Wilberforce Road, Cambridge, CB3 OWA, UK}}
~\\[0.15cm]
{$^2$~{\it Department of Mathematics, \\
King’s College London, Strand, WC2R 2LS, UK}}
~\\[0.15cm]
{$^3$~{\it Simons Center for Geometry and Physics, SUNY, Stony Brook, NY 11794, USA}}
~\\[0.15cm]
{$^4$~{\it Physique Th\'eorique et Math\'ematique and International Solvay Institutes
Universit\'e Libre de Bruxelles; C.P. 231, 1050 Brussels, Belgium}}

~\\[1.25cm]
			
\end{center}
\noindent We study various aspects of global symmetries in five-dimensional superconformal field theories. Whenever a supersymmetry-preserving relevant deformation is available, the infrared gauge theory description might exhibit a finite order mixed 't Hooft anomaly between a $1$-form symmetry and the instantonic symmetry. This anomaly constrains the flavor symmetry group acting faithfully on the SCFT and the consistency of certain RG flows.
As an additional example, we consider the instructive case of three-dimensional $\mathcal{N}=4$ SQED. Finally, we discuss the compatibility between conformal invariance and the presence of $1$-form and $2$-group global symmetries.

\vfill 
\begin{flushleft}
January 2022
\end{flushleft}
\end{titlepage}
%
		
	
\setcounter{tocdepth}{3}
\renewcommand{\cfttoctitlefont}{\large\bfseries}
\renewcommand{\cftsecaftersnum}{.}
\renewcommand{\cftsubsecaftersnum}{.}
\renewcommand{\cftsubsubsecaftersnum}{.}
\renewcommand{\cftdotsep}{6}
\renewcommand\contentsname{\centerline{Contents}}
	
\tableofcontents


\section{Introduction}

A fascinating question is whether non-abelian gauge theories in $d>4$ admit an ultraviolet fixed point. The study of these theories in $d=4+\epsilon$ reveals that a controlled UV fixed point does exist for $\epsilon \ll 1$ \cite{Peskin:1980ay}. Unfortunately, it is not currently known if such fixed point survives in $d=5$.\footnote{Recently there has been renewed interest in this subject  \cite{DeCesare:2021pfb,Florio:2021uoz}.}

To date, there exists a variety of string theory constructions pointing to the existence of $5d$ interacting superconformal field theories which should be thought of as isolated UV fixed points of the RG flow. Note that the unique five-dimensional superconformal algebra for which a stress tensor multiplet can exist is the $\cN=1$ algebra $\mf{f}(4)$, whose bosonic subalgebra is $\mf{so}(2,5)\times \mf{su}(2)_R$ \cite{Nahm:1977tg}, and its unitary superconformal multiplets have been classified \cite{Bhattacharya:2008zy, Buican:2016hpb, Cordova:2016emh}. These theories can be thus defined in purely field theoretical terms by an algebra of local operators obeying general consistency requirements from superconformal invariance. 

In this work we take this perspective to study examples of $5d$ SCFTs admitting a conserved (possibly non-abelian) global symmetry $G\zero_{CFT}$ with an associated flavor current multiplet. The only available supersymmetric relevant deformation of these theories resides at level two in this multiplet \cite{Cordova:2016xhm}. It is a (real) flavor mass deformation $M_{F}$, transforming in the adjoint representation of $G\zero_{CFT}$, and neutral under $\mf{su}(2)_R$. The RG flows triggered by activating $M_{F}$ does often give rise to an IR weakly-coupled description consisting of a $\cN=1$ $5d$ supersymmetric gauge theory.

Gauge theories in five dimensions admit a conserved $G\zero_{\textrm{ab},I}$ continuous Abelian symmetry. The charged objects are disorder operators that can be thought of as uplifts of (codimension-$4$) Yang--Mills instantons. They are also often referred to as instantonic particles. In a supersymmetric gauge theory, there are fermionic zero modes in the background of such instantonic configurations. Strikingly, the quantization of fermionic zero modes in an instanton background may give rise to additional conserved currents which combine with the original $G\zero_{\textrm{ab},I}$ conserved current (and potentially with other flavor currents if matter is also present) to form a non-abelian $G\zero_{\textrm{non-ab},I}$ conserved current \cite{Tachikawa:2015mha}. This is a non-perturbative enhancement of global symmetry. A similar phenomenon is observed in $3d$ supersymmetric gauge theories where the non-perturbative enhancement is due to monopole operators \cite{Borokhov:2002ib,Borokhov:2002cg, Bashkirov:2010kz}.\footnote{Of course, there can also be global symmetry enhancement due to local vector operators which become conserved at the fixed points. In four dimensions this is the only way that an enhancement can appear.}

Possibly the simplest example where the above symmetry enhancement is realized is the five-dimensional $\cN=1$ gauge theory consisting of a single $SU(2)$ vector multiplet. This should be thought of as the weakly-coupled description obtained by triggering a flavor mass deformation of a $5d$ SCFT with a $\mf{su}(2)_I$ conserved flavor current multiplet, and was first identified in \cite{Seiberg:1996bd, Morrison:1996xf, Douglas:1996xp}. This SCFT is typically referred to as $E_1$ SCFT or simply $\cT_{E_1}$. The basic idea is that  $SU(2)$ super Yang--Mills theory has a conserved $U(1)_I$ current and that additional fermionic zero modes in the background of an instanton generate the remaining $\mf{su}(2)_I$ broken currents multiplets. This picture is further confirmed by supersymmetric localization computations \cite{Kim:2012gu} and analysis of the instanton Hilbert series \cite{Cremonesi:2015lsa}.

In addition to its instantonic symmetry, a pure $SU(2)$ gauge theory also admits a $\bZ\one_2$ $1$-form symmetry, in the terminology of \cite{Gaiotto:2014kfa}. The charged object is a Wilson line in the fundamental representation of $SU(2)$. This theory has a mixed finite order 't Hooft anomaly between the $U(1)\zero_I$ and $\bZ\one_2$ encoded in the six-dimensional theory \cite{BenettiGenolini:2020doj}:
\be\label{schematically}
 A_6[\cA_I, \cB] = \exp \left( 2\pi \ii \frac{1}{4} \int_{\cY_6} c_1(\cA_I) \cup \cP(\cB) \right) \ec
\ee
where $c_1(\cA_I)$ is the first Chern class for the $U(1)\zero_I$ bundle whose background gauge field we denote by $\cA_I$, $\cP(\cB)$ is the Pontryagin square of the background gauge field $\cB$ for $\Z_2\one$, and $\partial \cY_6 = \cM_5$. 
In the supersymmetric context, one could consider the idea that $\bZ\one_2$ is not just an emergent IR symmetry of the pure $SU(2)$ gauge theory but rather a symmetry of the original $\cT_{E_1}$ SCFT. This is supported by recent string theory constructions in \cite{Morrison:2020ool, Albertini:2020mdx}. Then, assuming that $\bZ_2\one$ is not emergent, one has to find a supersymmetric RG flow which is consistent with the 't Hooft anomaly \eqref{schematically}. This requires that the non-abelian flavor symmetry group of $\cT_{E_1}$ is $SO(3)_I\zero$, a fact also emphasized in the recent paper \cite{Apruzzi:2021vcu}, and that the theory has a mixed 't Hooft anomaly between $SO(3)_I\zero$ and $\Z_2\one$ \cite{BenettiGenolini:2020doj}. In this paper we will examine further implications of this proposal.

We discuss the constraints imposed by conformal invariance in $d>2$ on theories with a set of genuine line operators charged under a $1$-form symmetry. We observe that, in this case, an unbroken $1$-form symmetry is not compatible with conformal invariance, while a spontaneously broken $1$-form symmetry is always compatible with conformal invariance. An important caveat is that our remarks do not apply to $1$-form symmetries that act trivially on the RG fixed point. This could happen at the end points of certain RG flows involving UV theories with genuine line operators charged under a $1$-form symmetry. Another example where our remarks do not apply is in theories with a decoupled sector that one could, for instance, obtain by considering a theory which is the tensor product of a CFT and a topological quantum field theory with a $1$-form symmetry. Clearly, in this case, there is no incompatibility with conformal invariance.\footnote{In this paper we restrict our considerations to the action of $1$-form symmetries on lines. It would be interesting to extend the scope of the investigation to include the action of $1$-form symmetries on higher-dimensional operators via symmetry fractionalization.\label{footnote:careful_action}}

In the case of $\cT_{E_1}$, we expect that the above caveat does not apply and there exists a set of $1/2$-BPS genuine line operators  charged under $\bZ\one_2$. Since these line operators are rigid \cite{Agmon:2020pde}, upon triggering a flavor mass deformation they can be described as supersymmetric Wilson line of the pure $SU(2)$ gauge theory. Thanks to supersymmetric localization and AdS/CFT one can then perform some tests on these class of line operators and verify that they indeed obey perimeter law (e.g. \cite{Assel:2012nf}). 

We also show how to match the anomaly \eqref{schematically} in the phase with spontaneously broken $\bZ\one_2$. This is obtained by constructing a $\bZ_2$ gauge theory coupled to $\cA_I$ and $\cB$.

Using the observations about spontaneously broken $1$-form symmetries in a conformal field theory, we also comment on the recent findings that the $\cT_{E_1}$ can have a $2$-group symmetry involving $\Z_2\one$ \cite{Apruzzi:2021vcu}. More generally, if we assume as above that the $1$-form symmetry does not act trivially on the CFT, we observe an obstruction in realizing a spontaneously broken $2$-group. The basic reason is that the allowed patterns of symmetry breaking patterns of a $2$-group are very restrictive \cite{Cordova:2018cvg, Benini:2018reh}. Similar constraints on $2$-group symmetry breaking patterns have also been exploited in different contexts \cite{Gukov:2020btk,Brennan:2020ehu}.

Another application of our analysis is in the context of supersymmetric RG flows of $5d$ SCFTs. It has been proposed by various authors that different-looking gauge theories can be obtained from inequivalent $G\zero$ flavor mass deformations of a single $5d$ SCFT, so that their  protected observables match \cite{Bergman:2013aca, Gaiotto:2015una}. By an abuse of language, this is often referred to as a ``UV duality." Here we propose a simple observation relying on the assumption that the $1$-form symmetry $G\one$ is not emergent (see also \cite{Gukov:2020btk}). If two different-looking IR gauge theories, with the same symmetries $G_{IR}\zero \times G\one$, have mixed anomalies \eqref{schematically} that do not match, they cannot arise from inequivalent $G_{IR}\zero \times G\one$-preserving deformations of the same UV fixed point. This can be thought of an IR constraint on the consistency of the proposed RG equivalences. 

Finally, it is quite interesting to notice that similar problems arise also in a simple $3d$ $\cN=4$ supersymmetric $U(1)$ gauge theory with $N_f=2$ hypermultiplet of charge $1$, which we discuss in section \ref{sec:Warm-up}. Even though this theory doesn't have a $1$-form symmetry, it does have a finite order mixed 't Hooft anomaly between the monopole $U(1)\zero_T$ symmetry and the faithful flavor symmetry group $SO(3)_F\zero$. It is widely expected that this theory flows to a superconformal $\cN=4$ theory in the IR with enhanced non-abelian topological symmetry. The mixed anomaly in this case is only consistent with an IR global symmetry group $SO(3)\zero_T$. In three dimensions, one can use mirror symmetry to track $SO(3)\zero_T$ in the dual theory. In the dual picture, this symmetry is simply the manifest flavor symmetry group. One could say that consistency with IR dualities implies that the original theory must have emergent non-abelian topological symmetry enhancement. Indeed, analogous ideas have been proven to be very effective in the recent study of non-supersymmetric $3d$ dualities.

\section{Symmetries of \texorpdfstring{$\cN=4$}{N=4} SQED in Three Dimensions}
\label{sec:Warm-up}

\subsection{Symmetry Enhancement}
\label{subsec:3dSymmEnhancement}

An illuminating example showcasing all the central themes in this paper is a model of three-dimensional $\cN=4$ gauge theory describing the interaction of a single Abelian vector multiplet $\cV$ with $N_f$ charged hypermultiplets $\cQ$.\footnote{In what follows we will always consider charge $1$ hypermultiplets and also $N_f \geq 2$.} This theory has received a lot of attention since the works \cite{Intriligator:1997pq,Kapustin:1999ha} and it is commonly referred to as $\cN = 4$ SQED.\footnote{We will sometimes use the notation SQED$_{N_f}$.} Using $\cN=2$ notation, the hypermultiplet consists of two chiral superfields $Q, \tilde Q$ of charge $1,-1$ respectively. The vector multiplet $\cV$, instead, can be decomposed in terms of a real vector superfield $V$, whose bottom component is a real scalar, and an adjoint chiral $\Phi$ whose bottom component is a complex scalar. Apart from the standard kinetic terms for $\cV$ and $\cQ$ (now expressible in terms of $\cN=2$ superfields), the action of $\cN=4$ SQED contains a superpotential term of the form
\be
\int \rd^3x \, \rd^4\theta \sum^{N_f}_{i=1} Q_i \Phi \tilde{Q}_i + \textrm{h.c}\ed
\ee 

The $R$-symmetry group of $3d$ $\cN=4$ theories is $G_R = SU(2)_C \times SU(2)_H$.\footnote{In this section we only consider ordinary symmetries, so we write $G$ for $G\zero$ without possible confusions.} The bottom components of $Q$ and $\tilde{Q}$ are singlets under $SU(2)_C$ while the bottom components of both $V$ and $\Phi$ transforms as singlets under $SU(2)_H$. 
These theories also exhibit interesting moduli spaces of vacua.  
The \emph{Coulomb branch}, denoted by $\cM_C$, is a complex manifold characterized by a non-zero expectation value of the scalar components in a $\cN=4$ vector superfield $\cV$. More precisely, $\cM_C$ is a hyper-K\"{a}hler manifold with K\"{a}hler forms rotating as a vector of $SU(2)_C$ and completely invariant under $SU(2)_H$. The \emph{Higgs branch} $\cM_H$ is also a hyper-K\"{a}hler manifold, now obtained by turning on an expectation value for the scalar components of $\cQ$. In contrast to the Coulomb branch, the Higgs branch is protected against quantum corrections. The action of $SU(2)_C$ on $\cM_H$ is trivial while its K\"{a}hler forms rotate as a vector of $SU(2)_H$. Note that a supersymmetric vacuum appearing at the intersection of $\cM_C$ and $\cM_H$ should necessarily be $SU(2)_C \times SU(2)_H$ invariant, and it is a candidate for an interacting critical point.

In addition to the $R$-symmetry symmetry group, $\cN=4$ SQED$_{N_f}$ also has a $SU(N_f)$ flavor symmetry group acting on the hypers $Q_i$ with $i= 1, \dots, N_f$. Notice also that a $\bZ_{N_f}$ center transformation $Q_i \to e^{\frac{2\pi \ii}{N_f}}Q_i$ coincides with a $U(1)$ gauge transformation. It is important to identify the global flavor symmetry group under which the gauge invariant operators transform faithfully which, in this case, is $PSU(N_f) = SU(N_f)/\bZ_{N_f}$. In order to keep track of this symmetry, we activate a background gauge field $\cB_F \in H^2(\cM_3, \bZ_{N_f})$, where $\cM_3$ is the spacetime.
We emphasize that this procedure does not modify in any way the original theory and it should just be thought of as studying a new observable in $\cN=4$ SQED$_{N_f}$. Compared to $SU(N_f)$, the group $PSU(N_f)$ allows for more choices of bundles $\cE$ characterized by a non-trivial Brauer characteristic class $w_2(\cE) \in H^2(\cM_3, \bZ_{N_f})$. Upon introducing $\cB_F$, the $PSU(N_f)$ flavor bundles in the path integral are constrained by fixing 
\be
\cB_F = w_2(\cE) \ed
\ee
Crucially, the identification of central flavor symmetries and gauge transformations relates the Brauer class and the gauge field, because it modifies the quantization condition
\be
\label{frac}
\int \left( \frac{F}{2\pi} - \frac{1}{N_f}\cB_F \right) \, \in \bZ \ec 
\ee
where $F = \rd A$ is the curvature of the $U(1)$ gauge field.
We thus see that considering a $PSU(N_f)$ background field $\cB_F$ induces charge fractionalization. 

Finally, there is another important global symmetry that we need to describe. Any three-dimensional $U(1)$ gauge theory has a conserved current given by
\be
\label{eq:3dTopologicalSymmetry}
J_T = \star \frac{F}{2\pi} \ec \qquad q_T= \int \star J_T \, \in \bZ \ed
\ee
This conserved current is associated to a topological symmetry. If $F$ is the curvature of a true $U(1)$ gauge bundle, then $q_T\in\Z$, and we denote the topological symmetry as $U(1)_T$. The local operators charged under $U(1)_T$ are disorder operators which change the boundary conditions of gauge fields at a point in spacetime. They are referred to as \emph{monopole operators}. In $\cN=4$ SQED, these become chiral operators which in $\cN=2$ language can be constructed explicitly in terms of the bottom components of $V$ and $\Phi$. On the Coulomb branch, the $U(1)_T$ is spontaneously broken and acts non-linearly as a shift symmetry for the dual photon. It is standard to introduce a background gauge field $\cA_T$ for $U(1)_T$ and minimally couple it to $J_T$ as
\be
\delta \scrL = \ii \cA_T \wedge \star J_T \ed
\ee
If there is a non-trivial background for $PSU(N_f)$ as in \eqref{frac}, the minimal coupling described above is no longer invariant under large gauge transformations for $U(1)_T$ \cite{Komargodski:2017dmc}. This signals a finite order mixed 't Hooft anomaly between $U(1)_T$ and $PSU(N_f)$ which can be conveniently described in terms of a four-dimensional inflow term
\be\label{an3d}
A_4[\cA_T, \cB_F] = \exp\left( - 2\pi \ii \frac{1}{N_f} \int_{\cY_4} \frac{\cF_T}{2\pi}  \, \cB_F \right)\ec
\ee
where $\partial \cY_4 = \cM_3$ and $\cF_T = \rd \cA_T$ locally.

Without any hypermultiplet mass or FI term for the vector multiplet, the only scale in $\cN=4$ SQED is the gauge coupling $g_{3d}^2$, and it is widely believed that this theory flows in the infrared to a superconformal fixed point lying at the intersection of $\cM_C$ and $\cM_H$. At the superconformal point, the monopoles transform in a superconformal multiplet: the bottom component has scaling dimension $q_R$ and transforms in a representation of $SU(2)_C$ of dimension $2q_R+1$, where \cite{Borokhov:2002cg}
\be
\label{Rmonopole}
q_R = \frac{1}{2} |q_T N_f| \ed
\ee

In the infrared $\cN=4$ superconformal theory, the topological current $J_T$ sits in a flavor current multiplet, and the bottom component of such superconformal multiplet is a Lorentz scalar of $R$-charge $1$. Looking at \eqref{Rmonopole}, we notice that something special happens for $N_f=2$: in this case, the basic monopole with topological charge $q_T = \pm 1$ has $R$-charge $q_R=1$, and thus it necessarily belongs to a new conserved flavor current multiplet \cite{Gaiotto:2008ak}. Since this current does not commute with the $U(1)_T$ current, monopole operators give rise to a phenomenon of IR global symmetry enhancement: SQED$_2$ with $U(1)_T$ symmetry flows to a fixed point with a larger non-abelian symmetry $\mf{su}(2)_T$.

Given this, we should consider the problem of anomaly matching for \eqref{an3d} when $N_f=2$. Clearly, all the symmetries involved in the UV mixed 't Hooft anomaly are preserved in the IR and one should be able to write a corresponding inflow term. However, if the enhanced topological global symmetry Lie algebra $\su(2)_T$ is promoted to a group $SU(2)_T$ we run into a problem, since there is no available degree-two characteristic class. It would then be impossible to write a nontrivial four-dimensional inflow term describing the anomaly in the IR.

The above difficulty disappears by identifying the topological symmetry group acting faithfully. A particularly clear way to understand point goes back to the work of \cite{Seiberg:1996nz}. By going on the Coulomb branch (where the topological symmetry acts on the monopole operators), one can study a non-vanishing amplitude with fermionic zero modes in the background of a BPS monopole operator and establish that such amplitude would lead to a non-anomalous contribution if and only if the faithful topological symmetry group acting on the monopole operators everywhere on $\cM_C$ is actually $SU(2)_T/\bZ_2 = SO(3)_T$.

Alternatively, one can look at the structure of the IR $\cN=4$ SCFT characterized by the superalgebra $Osp(4|4)$ where monopole operators have dimension $q_R$ and transform as irreducible $\mf{su}(2)_T$ representations of dimensions $2q_R+ 1$. A similar structure is also manifest in the Coulomb branch Hilbert series \cite{Cremonesi:2013lqa}. All gauge invariant operators charged under the topological symmetry transform in representations which are neutral under the $\bZ_2$ center symmetry, providing further support to the enhancement of the topological symmetry to $SO(3)_T$. 

This enhancement pattern of the topological symmetry solves the problem of 't Hooft anomaly matching. With the enhancement $SO(2)_T\to SO(3)_T$, we can identify the first Chern class of $SO(2)_T\cong U(1)_T$ with the second Stiefel--Whitney class of the $SO(3)_T$ background gauge bundle
\be
\label{embedding}
c_1(U(1)_T) = \frac{\cF_T}{2\pi} = {w_2(SO(3)_T)} \mod 2 \ed
\ee 
This leads to matching the anomaly \eqref{an3d} with 
\be
\label{an3dIR}
A^{\text{IR}}_4[\cB_T, \cB_F] = \exp\left( -2\pi \ii \frac{1}{2} \int_{\cY_4} \cB_T \, \cB_F \right) \ec
\ee
where $\cB_T$ is the background gauge field that is set equal to $w_2(SO(3)_T)$ (see also \cite{Gang:2018wek, Eckhard:2019jgg} for a related discussion).

\subsection{Mirror Symmetry}

Much of the attention devoted to $\cN=4$ SQED is due to the discovery of a three-dimensional IR duality known as mirror symmetry, which we now briefly review in the specific example of $\cN=4$ SQED$_{N_f}$ with $N_f>2$. We refer to SQED$_{N_f}$ as ``Theory A" and denote it by $\cT_A$. As discussed above, $\cT_A$ is characterized by the following UV faithful global symmetries
\be
\label{symmetriesA}
SO(3)_C \times SO(3)_H \times PSU(N_f)_{F} \times U(1)_T \ed
\ee
According to the original proposal of mirror symmetry, there is a ``Theory B"  $\cT_B$ that flows in the infrared to the same fixed point as $\cT_A$. This latter theory is $U(1)^{N_f}/U(1)_{\text{diag}}$ gauge theory with $N_f$ twisted hypermultiplets denoted in $\cN=2$ notation by $(q_i, \tilde{q_i}), i=1, \dots, N_f$ \cite{Intriligator:1996ex}.\footnote{Twisted hypermultiplets are analogous to hypermultiplets, but the role of $SU(2)_C$ and $SU(2)_H$ are exchanged.} Under the gauge group, the twisted hypermultiplets transform as
\beq
q_i \to \e^{\ii (\alpha_i - \alpha_{i-1})} q_i \, , \qquad \tilde{q_i} \to \e^{-\ii (\alpha_i - \alpha_{i-1})} \tilde{q_i} \, , 
\eeq
for all $i=1, \dots, N_f$ with $\alpha_0 = \alpha_{N_f}$. There is a flavor symmetry $U(1)_{\hat  F}$ acting as
\be
\label{eq:QuiverFlavorSymmetry}
q_i \to e^{\ii\beta} q_i\ec\qquad  \tilde{q}_i \to e^{-\ii\beta} \tilde{q}_i\ed
\ee
We denote by $J_{\hat  F}$ its associated Noether current. In addition, we have $N_f-1$ conserved topological symmetries, which we denote by $U(1)^{N_f-1}_{\hat T}$, with currents
\be
J_{\hat T_i} = \star \frac{F_i}{2\pi} \ec \qquad q_{\hat T_i}= \int \star J_{\hat T_i} \, \in \bZ\ec \qquad \sum J_{\hat{T}_i} = 0 \ec \qquad i = 1, \dots, N_f \ed
\ee

As in the previous case, it is important to identify the faithful global symmetry. First, we observe that the gauge-invariant operators are composite operators
\beq
q_1 q_2 \cdots q_{N_f} \, , \qquad \tilde{q}_1 \tilde{q}_2 \cdots \tilde{q}_{N_f} \, ,
\eeq
and these carry $U(1)_{\hat{F}}$ charge in multiples of $N_f$, which means that the global symmetry should be thought of as $U(1)_{\hat{F}}/\Z_{N_f}$.\footnote{Of course, there are also operators $q_i\tilde{q}_i$, but these are neutral under $U(1)_{\hat{F}}$.} Thus, we define a $\tilde{U(1)_{\hat{F}}} \cong U(1)_{\hat{F}}/\Z_{N_f}$ gauge field by rescaling with $N_f$ $\tilde{\cA_{\hat{F}}} \equiv N_f \cA_{\hat{F}}$. The fluxes of $\tilde{\cA_{\hat{F}}}$ are quantized in integer multiples of $2\pi$. Relatedly, notice that a $U(1)_{\hat{F}}$ transformation with parameter $\beta = 2\pi k/N_f$ with integer $k$ is just a gauge transformation. Therefore, we can identify a subgroup $\Z_{N_f} \subset U(1)_{\hat{F}}$ of flavor transformations with gauge transformations, which imposes a relation between the fluxes of the two:
\beq
\int \left( \frac{F_i}{2\pi} - \frac{1}{N_f} \frac{\tilde{\cF_{\hat{F}}}}{2\pi} \right) \in \Z \, \qquad i= 1, \dots, N_f-1 \, .
\eeq
That is, the background gauge bundle for the $\tilde{U(1)_{\hat{F}}}$ symmetry controls the fractional part of each $U(1)/\Z_{N_f}$ gauge bundle. In turn, the fractionalization of the latter implies that the topological charges would also be fractional. Theory $B$ is thus characterized by the following set of UV global symmetries
\be\label{symmetriesB}
SO(3)_{\hat C} \times SO(3)_{\hat H} \times \frac{U(1)_{\hat  F}}{\Z_{N_f}} \times U(1)^{N_f-1}_{\hat T} \ed
\ee
As already seen for Theory A, the quotients in the global symmetry group determine a non-trivial transformation of the partition function in presence of a non-trivial background for $U(1)_{\hat{F}}$. Namely, upon performing a large gauge transformation for the background gauge field $\cA_{\hat{T}_i}$, the variation of the partition function can be obtained by inflow from the four-dimensional theory
\beq
\label{an3db}
A_4[\cA_{\hat{T_i}}, \tilde{\cA_{\hat{F}}}] = \exp \left( - 2\pi \ii \frac{1}{N_f} \int_{\cY_4} \frac{\cF_{\hat{T_i}}}{2\pi} \, \frac{\tilde{\cF_{\hat{F}}}}{2\pi} \right) \, .
\eeq
This variation cannot be cancelled by a properly quantized local counterterm in the background gauge fields. Therefore, it represents a fractional contact term in the two-point function of the Abelian flavor currents, which are scheme-independent well-defined physical observables \cite{Closset:2012vp}, corresponding to a fractional Hall conductance. The fractionalization of the Hall conductance is tantamount to the fact that the global symmetry is the quotient $U(1)_{\hat{F}}/\Z_{N_f}$: indeed, in terms of the background gauge field $\cA_{\hat{F}}$ for the $N_f$-cover, the local mixed Chern--Simons counterterm cancelling the variation \eqref{an3db} is properly quantized (see a thorough discussion in terms of $U(1)$ symmetry fractionalization in \cite{Cheng:2022nji}).

The basic idea of $3d$ mirror symmetry in this example is that both $\cT_A$ and $\cT_B$ flow to a non-trivial RG fixed point in the infrared and that each of these fixed points can be described in terms of a dual description using the mirror symmetry duality map. Notice that, in contrast with other dualities (such as Seiberg's original four-dimensional case \cite{Seiberg:1994pq}), the global symmetries only need to match in the IR but not in the UV. Under the map, the Coulomb branch of the theory is exchanged with the Higgs branch of its mirror theory
\be\label{moduliswitch}
\cM_C \leftrightarrow \cM_{\hat H} \ec \qquad \cM_H \leftrightarrow \cM_{\hat C}\ec
\ee 
and thus the $R$-symmetries of both sides are exchanged. Moreover, electrically charged particles and Abrikosov--Nielsen--Olesen vortices are exchanged \cite{Aharony:1997bx}. Thus, flavor and topological symmetries should be related. More precisely, the $U(1)_T$ current in $\cT_A$, $J_T$, is mapped to the $U(1)_{\hat F}$ flavor symmetry current of $\cT_B$ by
\be
J_T \leftrightarrow \frac{1}{N_f} J_{\hat F} \, .
\ee
Notice that this is only consistent if the integral of $\star J_{\hat{F}}$ is an integer multiple of $N_f$, which is due to the fact that \eqref{eq:QuiverFlavorSymmetry} is really a $U(1)/\Z_{N_f}$ symmetry, but mirror symmetry exchanges the gauge-invariant monopoles in $\cT_A$ with the baryonic operators $q_1\cdots q_{N_f}$ in $\cT_B$. Thus, the relation between the background gauge fields is $
\cA_T \leftrightarrow \tilde{\cA_{\hat{F}}}$.

It is a clear prediction of mirror symmetry that the $U(1)_{\hat{T}}^{N_f-1}$ symmetry should enhance in the infrared to $PSU(N_f)_{\hat{T}}$. The currents corresponding to the maximal torus inside the $PSU(N_f)$ global flavor symmetry in $\cT_A$, which we denote by $J_{F_i}$ and are constrained by the fact that their sum must vanish, are mapped to the $N_f-1$ conserved topological currents $J_{\hat T_i}$ by
\be\label{jfmapping}
J_{F_i} \leftrightarrow J_{\hat T_i} \ed 
\ee
The remaining non-Abelian flavor currents in Theory A can be mapped looking at the representation theory structure of superconformal conserved flavor current multiplets in the $\cN=4$ SCFT (as done, for instance, in \cite{Bashkirov:2010kz}).

Corresponding to these mappings, mirror symmetry exchanges FI parameters and real masses, since both of them can be interpreted as flavor masses induced by weakly gauging the topological symmetry and flavor symmetry, respectively.

In light of this discussion, it is worth understanding how finite order 't Hooft anomalies such as \eqref{an3d} match across the duality. First, we consider the case with two hypermultiplets: SQED$_2$ is self-dual under mirror symmetry (since the Theory B described earlier is isomorphic to SQED$_2$). Indeed, observe that the expression \eqref{an3dIR} is  symmetric under the exchange of the background gauge fields for the flavor and topological symmetry. In fact, another way of arguing in favor of the enhancement pattern $U(1)_T\to SO(3)_T$ is based on mirror symmetry, since it should be exchanged with the $SO(3)_F$ faithful global symmetry.

For the case with $N_f >2$, instead, naively we cannot match the anomaly \eqref{an3d} along the RG flow, since the symmetries do not even match. However, focusing on $\cT_A$, suppose we add $\cN=4$ real masses to the hypermultiplets, breaking $PSU(N_f)_F$ to $U(1)^{N_f-1}_F/\Z_{N_f}$, and then turn on background gauge fields for the resulting symmetry. It is then possible to show that the Brauer class of the latter bundle is related to the curvature of the connections of each $U(1)_F/\Z_{N_f}$ factor by
\beq
\label{eq:FractionalPartConnectionCartan}
\int \left( \frac{\cF_{F_i}}{2\pi} - \frac{1}{N_f}w_2(\cE) \right) \in \Z \, \qquad i = 1, \dots, N_f-1 \, ,
\eeq
where $\cF_{F_i}$ is the curvature of the connection of any of the $U(1)/\Z_{N_f}$ bundles obtained in the maximal torus of $PSU(N_f)$. More rigorously, one should define the first Chern class of the $N_f$-th power of the line bundle, and then the relation would be
\beq
c_1 = w_2(\cE) \mod N_f \, ,
\eeq
which corresponds to \eqref{eq:FractionalPartConnectionCartan} upon taking the $N_f$-th root on the left-hand side. Once reduced in this way, the variation of the partition function of $\cT_A$ \eqref{an3d} looks like
\beq
A_4[\cA_T,\cA_{F_i}] = \exp \left( - 2\pi \ii \frac{1}{N_f} \int_{\cY_4} \frac{\cF_T}{2\pi} \, \frac{\cF_{F_i}}{2\pi} \right) \, .
\eeq
As discussed, at this stage this variation does not represent an 't Hooft anomaly, but rather a fractional contact term in the currents two-point function, which can still be matched across dualities, and indeed, using the identifications above, matches exactly \eqref{an3db}.

Another way of looking at this is that \eqref{eq:FractionalPartConnectionCartan} shows that at the infrared fixed point, where the flavor symmetry of $\cT_B$ is enhanced to $PSU(N_f)$, the fractional contact term \eqref{an3db} in the two-point function of Abelian currents becomes a fully-fledged anomaly between discrete symmetries matching \eqref{an3d}.

Before we proceed, it is useful to recapitulate the main features of this example:
\begin{itemize}
\item Whenever we describe global symmetries we should distinguish between the ultraviolet $G^{\text{UV}}$ and infrared $G^{\text{IR}}$ symmetries in a given theory. In particular, under RG flow there can be elements of $G^{\text{IR}}$ which are not present in $G^{\text{UV}}$ that appear as accidental symmetries or as a result of a quantum mechanical symmetry enhancement due to strong coupling effects. The second type of situation is sometimes referred to as a ``quantum (global) symmetry" \cite{Benini:2017dus}. In the cases just described, $G^{\text{IR}}_T=SO(3)_T$ in $\cT_A$ and $PSU(N_f)_{\hat{F}}$ in $\cT_B$ are examples of quantum symmetry.
\item If a given theory develops an IR quantum symmetry, we can only activate background fields (in the ultraviolet) for a subgroup of that symmetry $G^{\text{UV}} \subset G^{\text{IR}}$. However, if $G^{\text{UV}}$ participates in a 't Hooft anomaly we must be able to verify that there is a similar obstruction also in the infrared by embedding the anomaly for $G^{\text{UV}}$ into $G^{\text{IR}}$. For example,  in SQED$_2$ we have that $G^{\text{UV}}_T=U(1)_T$ has a mixed 't Hooft anomaly with $PSU(N_f)_F$ \eqref{an3d}. The theory flows to an IR superconformal fixed point with quantum symmetry $G^{\text{IR}}_T=SO(3)_T$, and we can find an embedding \eqref{embedding} to demonstrate that a similar 't Hooft anomaly is also present in the IR \eqref{an3dIR}. An alternative point of view on this aspect is that 't Hooft anomalies provide a new constraint on the allowed pattern of symmetry enhancement for $G^{\text{IR}}$. An analogous behavior holds for fractional contact terms in the two-point functions of currents.  
\item It is often possible to find two different theories providing alternate description of the same IR physics. An example of this phenomenon is 3$d$ $\cN=4$ mirror symmetry between $\cT_A$ and $\cT_B$. Dualities imply that a certain global symmetry which is manifest everywhere along the RG flow in one description might not be visible in the UV in the dual description and must necessarily emerge quantum mechanically. This is clear if we look at the UV global symmetries of both $\cT_A$ \eqref{symmetriesA} and $\cT_B$ \eqref{symmetriesB} in our example. Moreover, in order for the duality to hold, all 't Hooft anomalies and fractional contact terms in both descriptions must match. 
\end{itemize}

\section{Symmetries of \texorpdfstring{$5d$}{5d} SCFTs}
\label{5dSCFTs}

\subsection{Constraints from \texorpdfstring{$5d$}{5d} Superconformal Invariance}\label{zeroform5dscft}

Five-dimensional superconformal field theories can be defined abstractly in terms of an algebra of local operators obeying general consistency requirements imposed by unitarity, crossing symmetry and  superconformal invariance. The unitary superconformal multiplets that can appear in such theories have been fully classified \cite{Bhattacharya:2008zy, Buican:2016hpb, Cordova:2016emh}. 

The unique superconformal algebra in five dimensions is the exceptional algebra $\mf{f}(4)$ whose bosonic subalgebra is $\mf{so}(2,5)\times \mf{su}(2)_R$. 
Local operators of $5d$ SCFTs transform in unitary representations of this algebra, so their quantum numbers are labeled by
\be\nonumber
[j_1, j_2]^{(R)}_\Delta\ec
\ee
where $j_1, j_2\in \Z_{\geq 0}$ are Dynkin labels for the Lorentz algebra $\mf{so}(5)\cong \mf{sp}(2)$, $R\in \Z_{\geq 0}$ is a Dynkin label of the $R$-symmetry $\mf{su}(2)_R$ (its representations have dimensions $R+1$), and $\Delta\geq 0$ is the scaling dimension. 
Note that the supercharges $Q$ transform in the $[1,0]^{(1)}_{\frac{1}{2}}$, so there are eight of them, corresponding to five-dimensional $\cN=1$ supersymmetry.

A five-dimensional SCFT does not admit marginal deformations. Moreover, the only supersymmetry-preserving relevant deformation resides in the same supermultiplet as a conserved flavor current \cite{Cordova:2016xhm}. Because of this, here we solely focus on SCFTs admitting a flavor current multiplet. This is a protected supermultiplet containing the conserved current for a continuous $0$-form global symmetry $G\zero$ as top component.\footnote{The global $R$-symmetry has a conserved current belonging to the stress tensor supermultiplet and does not play any role in this discussion.}

The $5d$ flavor current multiplet, denoted by $C_1[0,0]_{3}^{(2)}$, has a superconformal primary that is a Lorentz scalar of conformal dimension $\Delta = 3$ transforming as a triplet of $\mf{su}(2)_R$. Acting on the primary with $Q$, we can readily obtain the remaining components of supermultiplet
\beq
\label{eq:CurrentSupermultiplet}
[0,0]_3^{(2)} \xrightarrow{ \ Q \ } [1,0]_{\frac{7}{2}}^{(1)} \xrightarrow{ \ Q \ } [0,0]^{(0)}_{4}\oplus[0,1]^{(0)}_{4} \, .
\eeq
The top component $[0,1]^{(0)}_{4}$ is a Lorentz vector of scaling dimension 4 which is a singlet under $R$-symmetry, and it represents the conserved current. For many purposes, it is more intuitive to name explicitly the components of the supermultiplet: the bottom component is denoted by $\mu^{a}_{(ij)}$, where $a$ is an index in the adjoint of $G\zero$, and $i,j=1,2$ are indices in the fundamental of $\mf{su}(2)_R$; the fermionic component $\psi^{a}_{\alpha,i}$ is a symplectic Majorana spinor transforming as a doublet of $\mf{su}(2)_R$. Finally, the top components are a real scalar operator $M_F^a$, and the $G\zero$ current $J^a_\mu$.

As we emphasized above, the unique supersymmetry-preserving relevant deformation of a $5d$ SCFT resides at level two in the flavor current multiplet. This is usually called a flavor mass deformation and takes the following form:
\be\label{flavormass}\delta \scrL = m_a M_F^a = m_a \Omega^{\alpha\beta}Q^i_\alpha Q^j_\beta \mu^a_{(ij)}\ec\ee 
where $\Omega^{\alpha\beta}$ is the $\mf{so}(5)\cong \mf{sp}(2)$ symplectic form. It is often very useful to think about this deformation as weakly gauging the flavor symmetry $G\zero$ by coupling the current supermultiplet to a background vector multiplet and giving a VEV to the scalars, thus breaking some of the flavor symmetry \cite{Seiberg:1993vc}. 

A given $5d$ SCFT can have a large space of supersymmetric flavor mass deformations depending on its global symmetry $G\zero$. If we explore this space we can easily find RG flows which end on a different fixed point with reduced global symmetry group (as in the $E_n$ series of \cite{Seiberg:1996bd, Morrison:1996xf, Douglas:1996xp}), or in weakly-coupled, generically non-abelian, five-dimensional $\cN=1$ supersymmetric gauge theories \cite{Seiberg:1996bd}. Since both $M_F^a$ and  $J_\mu^a$ belong to a protected multiplet, it is possible to track them reliably in the IR whenever a weakly-coupled gauge theory description is available.\footnote{A deformation by the bottom component $\mu^a_{(ij)}$ instead breaks supersymmetry. Elucidating the resulting IR physics in this case is still an open problem, see \cite{BenettiGenolini:2019zth, Bertolini:2021cew} for some recent work.} In particular, for each simply connected factor in the IR gauge group there is a $U(1)_I^{(0)}$ symmetry with topological current 
\beq
J_{I,s} = \frac{1}{4h^\vee} \star \Tr_{\rm adj} \left( \frac{F^{(s)}}{2\pi} \wedge \frac{F^{(s)}}{2\pi} \right)\ec 
\eeq
descending from a particular flavor current of the SCFT. Here $F^{(s)}$ denotes the curvature of the simple gauge group $G^{(s)}_{\rm gauge}$, the trace is in the adjoint representation and $h^\vee$ is the dual Coxeter number of $\mf{g}^{(s)}_{\rm gauge}$. 

The current $J_I$ bears close similarities with the three-dimensional topological current described around \eqref{eq:3dTopologicalSymmetry}. Both currents are conserved thanks to the Bianchi identity, and the local operators charged under these symmetry can be interpreted in the weakly-coupled description as disorder operators. In five dimensions these operators should be understood as the uplift of four-dimensional Yang--Mills instantons. These are commonly referred to as \emph{instanton operators} and their role is crucial to understand the structure of non-abelian flavor symmetry currents in $5d$ SCFTs \cite{Lambert:2014jna, Tachikawa:2015mha}.

\subsection{Remarks on Higher-Form Symmetries and Conformal Invariance}\label{1form5dscft}

In order to completely characterize a quantum field theory one should consider various generalization of the notion of global symmetry. A particularly important example that will be examined here is the notion of generalized $p$-form global symmetries $G\gp$ which are tied to the existence of a topological operator $U_g(\Sigma_{d-(p+1)})$ of codimension $p+1$, where $\Sigma_{d-(p+1)}$ is a closed manifold and $g$ is an element of the symmetry group $G\gp$.\footnote{In Euclidean signature, the operator $U_g(\Sigma_{d-(p+1)})$ is frequently referred to as ``symmetry defect."} The objects charged under $G\gp$ are $p$-dimensional, and if $p \geq 1$, the group $G\gp$ must be Abelian.

If a higher-form symmetry acts faithfully on a given quantum field theory, we can identify an order parameter for such symmetry and discuss a generalization of the notion of spontaneous symmetry breaking \cite{Gaiotto:2014kfa, Lake:2018dqm, Hofman:2018lfz}.
Focusing on $1$-form symmetries, the order parameter is a line operator.
We say that a $1$-form symmetry is \emph{unbroken} if a charged line operator obeys an area law. This means that the vacuum expectation value of the charged line operator on large circular loop falls off exponentially with the surface area enclosed in the loop. Conversely, if the vacuum expectation value of a line operator of large size obeys a perimeter law, the symmetry is spontaneously broken, as line operators on smooth loops undergo multiplicative renormalization that removes divergences proportional to the length of the loop.

An important consequence of the previous definition is that an unbroken $1$-form symmetry is not allowed in an interacting conformal field theory in $d\geq 3$, as there is no possibility for charged line operators to obey an area law.\footnote{Note that in three dimensions a continuous $1$-form symmetry cannot be spontaneously broken, because of the Mermin--Wagner--Coleman theorem for $p$-form symmetries. However, unitarity constraints in $3d$ CFTs do not allow for a continuous $1$-form symmetry \cite{Lee:2021obi}.} Instead, charged line operators obeying a perimeter law are always compatible with conformal invariance. This shows that if any (non-decoupled) 1-form symmetries exist in a conformal field theory, they must be spontaneously broken.\footnote{A famous example being four-dimensional $\cN=4$ $SU(N)$ super Yang--Mills theory.} In this argument we are assuming that there exist line operators charged under the 1-form symmetry. There are cases when the 1-form symmetry acts only on operators of lower codimension, and it would be interesting to investigate how the considerations in this paper extend to those theories (as already mentioned in footnote \ref{footnote:careful_action}).

The constraints of superconformal invariance impose further restrictions on the type of $G\one$ that can appear in five-dimensions. If $G\one$ is continuous, i.e. it is a $U(1)\one$ symmetry, there should be an associated conserved two-form current $J_{\mu\nu}$ residing in a conserved current multiplet. However, a close inspection of \cite{Cordova:2016emh} reveals that in any interacting $5d$ SCFTs there are no superconformal multiplets admitting a conserved current with quantum numbers $[2,0]_3$.
Therefore, $5d$ SCFTs cannot enjoy continuous $1$-form symmetries (see also \cite{Lee:2021obi}).

If a $5d$ SCFT has a $1$-form symmetry, it should thus be finite and spontaneously broken. The objects charged under the $1$-form symmetry are conformal line defects. In order to have analytic control, we focus on those that are also supersymmetric, which have been fully classified in \cite{Agmon:2020pde}. In any interacting five-dimensional SCFT there is a unique class of superconformal line defects: they are $1/2$-BPS and preserve the $1d$ superconformal algebra $D(2,1;2;0)\times \mathfrak{su}(2)_{\textrm{right}} \subset \mf{f}(4;2)$ (the relevant real form of $\mf{f}(4)$). The preserved bosonic subalgebra is given by $\mf{so}(2,1)_{\textrm{conf}}\times \mf{so}(4)_{\textrm{rot}} \times \mf{su}(2)_R$ where $\mf{so}(2,1)_{\textrm{conf}}$ is the $1d$ conformal algebra and the transverse rotations symmetry algebra is further decomposed into $\mathfrak{so}(4)_{\textrm{rot}} \simeq \mathfrak{su}(2)_{\textrm{left}} \times \mathfrak{su}(2)_{\textrm{right}}$.  In five dimensions, the set of broken charges induced by a superconformal line defect generate a broken supercurrent multiplet known as displacement multiplet $B_1([0]\one_{\frac{3}{2}},\mathbf{2})$.\footnote{A conformal line defect modifies the Ward identity for the bulk stress-tensor by a contact term \cite{Billo:2016cpy}:
\be
\partial_\mu T^{\mu \perp}(x)= -\delta(x_\perp)\mathsf{D^\perp}(x_\parallel)\ec
\ee
where we split $x=(x_\parallel,x_\perp)$ and $x_\parallel$ is the longitudinal coordinate along the line while $x_\perp$ denotes collectively the transverse coordinates. $\mathsf{D}^\perp$ is the so called displacement operator. In a $5d$ SCFT the displacement operator resides in the displacement supermultiplet $B_1([0]\one_{\frac{3}{2}},\mathbf{2})$ as top component. Here we label the unitary irreducible representations by $([j]^{(R)},\bf{j'+1})$ where $j,R,j' \in \bZ_{\geq 0}$ are Dynkin labels for $\mf{su}(2)_{\{\textrm{left},R,\textrm{right}\}}$.
} This is the only allowed option for line defects preserving both the transverse rotation group $\mathfrak{so}(4)_{\textrm{rot}}$ and $\mf{su}(2)_R$.\footnote{There also exists a set of ``minimal" lines, preserving a $\mf{osp}(1|2;\bR)$ superconformal algebra and a single real supercharge $Q$. These lines must also break
completely the $\mf{so}(4)_{\textrm{rot}}$ transverse rotations. For this reason we will not further consider them here.} The $1/2$-BPS superconformal line defects are also neutral under the ordinary global symmetry group $G\zero$ (whenever it is present). Under a bulk RG flow triggered by a flavor mass deformation as described in section \ref{zeroform5dscft}, and giving rise to a weakly-coupled IR description, we expect to be able to describe such line defects directly in the $5d$ supersymmetric gauge theory.

So far our discussion has focused on the possibility of identifying a class of supersymmetric line defects charged under a finite $1$-form symmetry. However, superconformal invariance is also compatible with the existence of a \emph{decoupled} finite $1$-form symmetry. By this we mean that it might be possible to identify a finite one-form symmetry $G\one$ together with an associated codimension-$2$ topological surface operator $U_g(\Sigma_{d-2})$ without any available line operators charged under it. One example where such phenomenon could arise is at the endpoints of certain RG flows of theories with a one-form symmetry, where the original line operators decouple \cite{Cordova:2018cvg}. Another example would be a tensor product of a CFT and a TQFT with a $1$-form symmetry. In these events the associated $G\one$ symmetry is automatically unbroken. An important aspect is that, even when a global symmetry is decoupled, it is still possible to activate background fields for such symmetry and study the response of the theory under background gauge transformations.

\subsection{Global Symmetry Enhancement of \texorpdfstring{$E_1$}{E1} SCFT}
\label{subsec:5dAnomalyMatching}

A useful case study for understanding the properties of $5d$ SCFTs is the so-called $E_1$ theory $\cT_{E_1}$. As the name suggests, this theory admits a flavor current supermultiplet \eqref{eq:CurrentSupermultiplet} whose non-abelian algebra is $\mf{su}(2)_I$. A common belief is that  activating a real flavor mass deformation \eqref{flavormass} leads to an IR weakly-coupled description given by a $5d$ $\cN=1$ supersymmetric Yang--Mills theory with gauge group $G= SU(2)$ and (dimensionful) gauge coupling inversely proportional to said flavor mass. In addition to the $SU(2)_R^{(0)}$ $R$-symmetry, which is always preserved by this type of RG flow, the global symmetries of $5d$ $SU(2)$ pure supersymmetric Yang--Mills are
\beq
\label{eq:SU2SYMSymm}
U(1)_I^{(0)}\times \Z_2^{(1)}\ed
\eeq
The first factor is an instantonic $U(1)\zero_I$ global symmetry, with conserved current
\beq
J_I = \frac{1}{8\pi^2}\star \Tr F\wedge F \ec
\eeq
descending from the triplet of conserved currents residing in the flavor current multiplet of the UV $E_1$ theory. The $U(1)\zero_I$ charged states are uplifts of four-dimensional Yang–Mills instantons which appear as ordinary massive particles in five dimensions with a mass proportional to the inverse IR gauge coupling $g^{-2}_5$.

The second factor in \eqref{eq:SU2SYMSymm} is a $\bZ\one_2$ symmetry associated to the center of the gauge group $SU(2)$, which acts on $1/2$-BPS fundamental Wilson loops described by
\be\label{BPSWilson}
W_F =\tr_F {\rm P} \exp\left( \ii \int A_\mu \dot{\gamma}^\mu + |\dot{\gamma}|\phi\right)\ec
\ee
where $\gamma$ is a straight line and $\phi$ is the real scalar component of a $5d$ vector multiplet.

As described in \cite{BenettiGenolini:2020doj}, the instantonic and center symmetries have a mixed 't Hooft anomaly, whose derivation is analogous to the three-dimensional anomaly \eqref{an3d} (a recent derivation using string theory methods is given in \cite{Apruzzi:2021nmk}). Again, we turn on a background gauge field for $U(1)_I^{(0)}$ as
\beq
\label{eq:MinimalCoupling5d}
\delta\ms{L} = \ii \cA_I \wedge \star J_I \, ,
\eeq
and also a background $\Z_2$ gauge field $\cB \in H^2(\cM_5,\Z_2)$ for the center symmetry.\footnote{Throughout the paper we assume that $\cM_5$ is spin.} This corresponds to including in the path integral additional $SO(3)$ bundles with second Stiefel--Whitney class $w_2 = \cB$. The integral of $\star J_I$ for these bundles (their instantonic number), though, is not integer, and the fractional part is controlled by the Stiefel--Whitney class as
\beq
\int \left( \frac{1}{8\pi^2}\Tr F \wedge F + \frac{1}{4}\cP(w_2) \right) \in \Z \, , 
\eeq
where $\cP:H^2(\cM_5, \Z_2)\times H^2(\cM_5, \Z_2) \to H^4(\cM_5, \Z_4)$ is the Pontryagin square. Therefore, we conclude that the minimal coupling $\delta \ms{L}$ is not anymore invariant under a large gauge transformation of $U(1)_I^{(0)}$: under $\cA_I\to \cA_I + \lambda^{(1)}$ with winding $2\pi\ell$, the partition function varies by
\beq
\label{eq:Variation5dPF}
Z[\cA_I, \cB] \to Z[\cA_I,\cB] \, \exp \left( - 2\pi \ii \ell \frac{1}{4} \int \cP(\cB) \right) \, .
\eeq
The mixed 't Hooft anomaly can be written in terms of a six-dimensional theory with partition function
\beq
\label{an5d}
A_6[\cA_I,\cB] = \exp \left( 2\pi \ii \frac{1}{4} \int_{\cY_6} \frac{\rd \cA_I}{2\pi} \, \cP(\cB) \right) \, ,
\eeq
where $\partial\cY_6 = \cM_5$. An analogous anomaly may also be found in super-Yang--Mills theories with different gauge groups and potentially even matter, as pointed out in \cite{BenettiGenolini:2020doj} and explained in Appendix \ref{app:OtherClassicalGroups} (see also \cite{Apruzzi:2021vcu}).

We shall now assume that $\cT_{E_1}$ admits a set of genuine $1/2$-BPS superconformal line defects, introduced in section \ref{1form5dscft}, charged under $\Z_2\one$. An important property of this class of superconformal line defects is that they are rigid, meaning that they do not admit couplings to exactly marginal parameters on their worldvolume \cite{Agmon:2020pde}.\footnote{To describe a supersymmetric deformation of a given superconformal line defect (in this context referred to as a  $1d$ defect conformal field theory) by a local operator $\cO$ living on its worldvolume $\gamma$, one should consider
\be S_{\textrm{DCFT}}\to S_{\textrm{DCFT}} + g_{\cO}\int_\gamma \cO(x_\parallel)\ed\ee However, the only available supersymmetry preserving deformations of  $1/2$-BPS superconformal line defects in $5d$ have $\Delta_{\cO}>1$.} To our knowledge, a detailed study of the behavior of generic $5d$ superconformal line defects under RG flows does not currently exist. A common belief in the literature, supported by supersymmetric localization and AdS/CFT correspondence, is that the superconformal lines of a UV fixed point are mapped, in the weakly-coupled IR SYM, to the supersymmetric Wilson lines \eqref{BPSWilson}, see e.g. \cite{Assel:2012nf}. Moreover, in this particular case, various recent string theory analysis \cite{Morrison:2020ool, Albertini:2020mdx}, have also supported the idea that $\cT_{E_1}$ admits a non-trivial $\Z_2\one$ symmetry.

Assuming that $\Z_2\one$ is not an emergent symmetry of the IR gauge theory phase, it makes sense to ask if there exists an anomaly involving such symmetry directly in the UV description. By 't Hooft anomaly matching, upon activating a flavor mass deformation as in \eqref{flavormass}, this putative anomaly would match the anomaly computed in the IR $SU(2)$ SYM theory description \eqref{an5d}. Notice that we cannot identify $c_1(U(1)\zero_I)$ with any characteristic class in $SU(2)_I\zero$, since there are no non-trivial degree-two characteristic classes. So, 't Hooft anomaly matching is only compatible with a symmetry breaking pattern $SO(3)_I\zero\to U(1)_I\zero$.
Therefore, we propose that the faithful global symmetry group of $\cT_{E_1}$ should be $G_{I}^{UV,}{\zero}= SO(3)\zero_I$ and that the anomaly of the theory is given by
\beq\label{anE1}
A^{UV}_6[\cB_I, \cB] = \exp \left( 2\pi \ii \frac{1}{4} \int_{\cY_6} \cB_I \, \cP(\cB) \right) \, ,
\eeq
where $\cB_I = w_2(SO(3)_I\zero)$. This conclusion, which was proposed in \cite{BenettiGenolini:2020doj}, is consistent with the analysis of the representations appearing in the superconformal index \cite{Bashkirov:2012re, Kim:2012gu}, the Higgs branch Hilbert series \cite{Cremonesi:2015lsa}, and the recent string theory analysis of \cite{Apruzzi:2021vcu}.

The discussion in five dimensions parallels that of section \ref{subsec:3dSymmEnhancement}, where we have used the same anomaly matching argument to argue that the IR (quantum) global symmetry of $3d$ SQED$_2$ is $SO(3)\zero_T$. 

\subsection{Anomaly Matching}\label{spontbreaking}

The mixed 't Hooft anomaly presented in section \ref{subsec:5dAnomalyMatching} leads to a number of dynamical implications. 

Let us first follow the anomaly along the Coulomb branch of the supersymmetric gauge theory. This is a \emph{real} moduli space of vacua parametrized by a non-zero expectation value for the bottom component $\phi$ of the five-dimensional vector multiplet.
The Coulomb branch of $SU(2)$ pure super Yang--Mills theory is simply $\R_+$, the corresponding low-energy effective theory is just a supersymmetric $U(1)$ gauge theory whose dynamics is governed by a real prepotential \cite{Intriligator:1997pq}. Note that both $U(1)\zero_I$ and $SU(2)_R$ symmetries are unbroken on the Coulomb branch.

An Abelian gauge theory in five-dimensions has emergent continuous higher-form symmetries, in the terminology of section \ref{1form5dscft}, $U(1)_e^{(1)}\times U(1)_m^{(2)}$. Since the theory is deconfined, the electric $U(1)_e^{(1)}$ must be spontaneously broken. Recall that when the non-abelian $SU(2)$ gauge symmetry is restored, the UV theory enjoys a $\bZ\one_2$ symmetry which is embedded into the emergent $U(1)_e\one$. In particular, the spontaneously broken $U(1)_e\one$ is realized non-linearly by a shift of the dynamical Abelian gauge field $a$:
\beq
\label{eq:TransformationAbelianGaugeCoulomb}
a \to a + \pi \, \epsilon \, , 
\eeq
where $\epsilon$ is a flat $\Z_2$ connection. 
It is then easy to see how the low-energy Abelian gauge theory on the Coulomb branch reproduces the anomaly \eqref{eq:Variation5dPF}. From the transformation law \eqref{eq:TransformationAbelianGaugeCoulomb}, we see that activating a background $\Z_2$ gauge field $\cB$ for $\bZ_2\one$ in the $SU(2)$ gauge theory corresponds to shifting the Abelian field strength $f$ by $f - \pi \cB$. The minimal coupling of $U(1)_I\zero$ then becomes
\beq
 \delta \ms{L} = \frac{\ii}{4\pi^2} \cA_I \wedge (f-\pi \cB) \wedge (f- \pi \cB) \, .
\eeq
Notice that upon changing the background $\cA_I$ by a large gauge transformation with winding $2\pi\ell$, we again find that the change of the partition function is equal to \eqref{eq:Variation5dPF}, thus reproducing the anomaly. 

Moving to the non-abelian gauge theory phase, pure $SU(2)$ supersymmetric Yang--Mills theory in five dimensions is an example of deconfined gauge theory with spontaneously broken $\bZ_2\one$ symmetry. Recall that a phase characterized by spontaneously broken finite higher-form symmetry includes a corresponding TQFT \cite{Gaiotto:2014kfa}. In this case, we describe such deconfined phase by a TQFT presented as a $\bZ_2$ gauge theory coupled to background fields with mixed 't Hooft anomaly \eqref{an5d}. To do so, we adopt a general method put forward in \cite{Hsin:2019fhf} that is analogous to the idea that Wess--Zumino--Witten terms match spontaneously broken continuous ordinary symmetries participating in 't Hooft anomalies.

The basic observation behind describing the spontaneous breaking of $\Z\one_2$ with a $\Z_2$ gauge theory is that this theory too has a $\Z_2\one$ symmetry, and the line operator charged under it is deconfined. The coupling of the TQFT to the $SU(2)$ SYM then consists in the identification of the line in the former with the fundamental Wilson line in the latter.

Operationally, we introduce a $\Z_2$ gauge field $u\in H^1(\cM_5, \Z_2)$, and we couple it to a background gauge field for the $\Z_2\one$ by modifying the closure condition to $\delta u=\cB$. Upon performing a gauge transformation of the background $\cB \to \cB + \delta \lambda$ (with a $\Z_2$ $1$-cochain $\lambda$), the constraint is satisfied by having the $\Z_2$ gauge field transforming non-linearly as $u\to u + \lambda$ with the same $\lambda$. This ultimately shows that $u$ describes a Goldstone mode for the broken symmetry, and the deconfined line in the $\Z_2$ gauge theory transforms by
\beq
\exp \left( 2\pi \ii \frac{1}{2} \int u \right) \to \exp \left( 2\pi \ii \frac{1}{2} \int u \right) (-1)^{\int \lambda}\ed
\eeq
To construct the $\bZ_2$ TQFT mentioned above we first compute the variation of the anomaly theory \eqref{an5d} 
\beq
A_6[\cA_I, \cB + \delta \lambda] = A_6[\cA_I, \cB] \, \exp \left( 2\pi \ii \frac{1}{4}\int_{\cM_5} \frac{\rd \cA_I}{2\pi} \, \left( \lambda \cup \delta \lambda + 2\lambda \cup \cB + \delta \lambda \cup_1 \cB \right) \right)\ec
\eeq
under a background gauge transformation for $\cB$.\footnote{Here we have used a relation that follows from the definition of the Pontryagin square: if $f,g\in C^2(X;\Z)$ are integer lifts of elements in $H^2(X;\Z_2)$, and $\cP(f) = f \cup f - f \cup_1 \delta f$, then
\[
\cP(f+g) = \cP(f) + \cP(g) + 2 f \cup g - \delta ( g \cup_1 f) \mod 4 \, ,
\]
where $\cup_1$ is higher cup product introduced by Steenrod, a bilinear operation of degree $-1$.}

This variation gives us the action of the $5d$ $\bZ_2$ TQFT that is coupled to the original $SU(2)$ supersymmetric Yang–Mills theory. In addition to the standard kinetic term for the $\Z_2$ gauge field $u$, we introduce the constraint $\delta u = \cB$, and the following coupling
\beq
\label{5dtqft}
S_{\rm c}= - 2\pi\ii \frac{1}{4} \int_{\cM_5}\frac{\rd \cA_I}{2\pi} \left( u \cup \delta u - \delta u \cup_1 \cB \right) \, .
\eeq
The constraint $\delta u = \cB$ enforces that $\Z_2\one$ acts on the lines of the $\Z_2$ gauge theory, and the coupling $S_{\rm c}$ above guarantees that the transformations $u\to u + \lambda$ and $\cB \to \cB + \delta \lambda$ on the TQFT reproduce the anomalous transformation of the original theory.

We stress that all the above manipulations are performed by looking solely at the transformation properties of the background field $\cB$. For this reason we can offer some comments regarding the $E_1$ fixed point. As we discussed in section \ref{1form5dscft}, conformal invariance is not compatible with an unbroken $\bZ\one_2$ symmetry. We have also seen that if such symmetry persists at the $E_1$ fixed point it should participate in a mixed 't Hooft anomaly with the enhanced instantonic global symmetry $SO(3)_I\zero$ \eqref{anE1}. The most compelling scenario, supported by evidence borne out of localization and holography \cite{Brandhuber:1999np, Assel:2012nf}, is that the $\bZ\one_2$ is spontaneously broken also at the fixed point. It is reassuring that a $\bZ_2$ TQFT analogous to \eqref{5dtqft} can be easily realized for the anomaly \eqref{anE1} using the strategy outlined in this section.

\subsection{Comments on \texorpdfstring{$2$}{2}-Group Global Symmetries}
\label{sec:2Groups}

Whenever a quantum field theory admits both a $0$-form $G\zero$ and $1$-form global symmetry $G\one$, it is often possible to find them intertwined into a new symmetry structure known as $2$-group global symmetry. In a $2$-group, the $0$-form and $1$-form symmetries combine non-trivially, as can be seen by the existence of junctions between the corresponding defects, which imply that the individual global symmetries participating in a $2$-group do not factorize. Here, we will not provide a detailed overview of $2$-group global symmetries but instead only highlight some of their basic properties that can be found in the literature, including \cite{Kapustin:2013uxa, Sharpe:2015mja, Tachikawa:2017gyf, Benini:2018reh}.

Since $5d$ SCFTs do not admit continuous $1$-form symmetries, we will only discuss $2$-group global symmetries involving finite $G\one$ symmetries. We will also assume that $G\zero$ is a compact (possibly non-abelian) Lie group. $2$-groups are defined by the following set of data:
\be
\mathbb{G}= (G\zero, G\one, \rho, [\beta])\ec
\ee
where $\rho$ is a homomorphism $\rho : G\zero \to \textrm{Aut}(G\one)$, and $[\beta]\in H^3_\rho(BG\zero, G\one)$ is a Postnikov class. Thus, $H^3(BG\zero,G\one)\neq 0$ is a necessary condition to have a 2-group. In the following, we shall restrict to trivial $\rho$, but the arguments may be generalized.

Concepts that are familiar in the study of ordinary global symmetries are also available for $2$-groups global symmetries. To describe some of them, we first activate background gauge fields $\cA$ for $G\zero$ and $\cB$ for $G\one$. In a theory with a $\mathbb{G}$ symmetry group, the junction of four $G\zero$ symmetry defects creates a flux for $G\one$ measured by the Postnikov class $[\beta]$. This is summarized in the general equation 
\be\label{2grpbkgd}
\delta \cB = \cA^*\beta\ec
\ee
which states that the background field $\cB$ is no longer flat but has differential fixed in terms of $\beta$ (a representative of the Postnikov class $[\beta]$). Here we pullback $\beta$ using $\cA$, viewed as a homotopy class of maps from spacetime to $BG\zero$, to obtain a $3$-cochain in spacetime taking values in $G\one$. Notice that $\beta$ cannot be described using only non-dynamical fields, but rather is a non-trivial operator of the theory. This is why one cannot interpret as an anomalous phase the appearance of $\beta$ as one changes the topology of the junction of four defects.

Concretely, at the level of infinitesimal background gauge transformations, the failure of $\cB$ to be closed \eqref{2grpbkgd} means that whilst $\cA$ transforms to $\cA'$ with $\lambda_\cA$ (be it a function valued in $\mf{g}\zero$ or a $0$-cochain valued in $G\zero$), $\cB$ transforms as
\beq
\label{eq:ChangeBackgroundB}
\cB \to \cB + \delta \lambda_\cB + \zeta(\lambda_\cA, \cA) \, ,
\eeq
where $\lambda_\cB$ is a $1$-cochain valued in $G\one$ (finite) and $\zeta(\lambda_\cA, \cA)$ is a $2$-cochain that satisfies
\beq
\label{eq:DefinitionZeta}
\delta \zeta(\lambda_\cA, \cA) = \cA'^*\beta - \cA^*\beta \, .
\eeq
Equation \eqref{eq:DefinitionZeta} may always be solved, but it defines $\zeta(\lambda_\cA, \cA)$ up to an exact piece, which can be compensated by changing $\lambda_\cB$.\footnote{Here we focus on the case of finite $G\one$, but it may be useful for the interested reader to compare with $G\one\cong U(1)$. In this case, if $G\zero$ is connected and simply connected, we have $H^3(BG\zero, U(1)\one)\cong H^4(BG\zero, \Z)\cong \Z$. When pulled back to $\cM_d$ using the $G\zero$ connection $\cA$, the generator of $H^4(BG\zero, \Z)$ is a multiple of the instanton number
\[
\left[ \frac{1}{4\pi}\tr \cF \wedge \cF \right] \, , 
\]
where $\cF$ is the curvature of $\cA$, so that the Postnikov class is related to the Chern--Simons form
\[
\cA^*\beta = \frac{\kappa}{4\pi}\tr\left( \cA \wedge \rd \cA + \frac{2}{3}\cA^3 \right) \, , \qquad \kappa \in \Z .
\]
Equation \eqref{eq:DefinitionZeta} may then be interpreted as a descent equation, and $\zeta$ is related to the primitive of the gauge-transformed Chern--Simons $3$-form
\[
\zeta(\lambda_\cA, \cA) = \frac{\kappa}{4\pi}\tr ( \lambda_\cA \rd \cA ) \, ,
\]
which is well-known to be only defined up to exact pieces. 
}

In a $\mathbb{G}$-symmetric theory it is always possible to introduce background fields satisfying \eqref{2grpbkgd}. This leads to a number of immediate consequences:
\begin{itemize}
    \item If $[\beta]$ is non-trivial, it is impossible to  couple a $\mathbb G$-symmetric theory uniquely to a background gauge field for $G\zero$. This is because the existence of a $0$-form symmetry background also introduces a non-trivial background for the $1$-form symmetry as seen from \eqref{2grpbkgd}. An intuitive way to phrase this statement is that the $0$-form symmetry group is never a good subgroup of a $2$-group global symmetry.
    \item Without activating any background for $G\zero$, it is always possible to only consider a background for $G\one$ and discuss its gauging. In other words, the $1$-form symmetry group is always a good subgroup of a $2$-group global symmetry.
    \item Finally, if $[\beta]=0$ the $2$-group $\mathbb G$ is trivial and \eqref{2grpbkgd} reduces to the flatness condition for $\cB$. A trivial $2$-group is nothing else than the direct product of $G\zero$ and $G\one$ symmetry groups.
\end{itemize}
Global symmetries can be unbroken or spontaneously broken by the vacuum. The allowed patterns of spontaneous symmetry breaking for a $2$-group $\mathbb G$ are:
\be\label{2grpscenarios}
\mathbb{G}\to G\one\qquad\textrm{or}\qquad \mathbb{G}\to \textrm{nothing}\ec
\ee
Intuitively, this follows from the general principle that $G\zero$ is not a good subgroup of a $2$-group. More concretely, in a broken symmetry phase we may still couple to the original background gauge fields participating in a $2$-group, but \eqref{2grpbkgd} and \eqref{eq:ChangeBackgroundB} show that it is not possible to couple to $G\zero$ without coupling to $G\one$. Because of this, a theory for a spontaneous symmetry breaking pattern that violates \eqref{2grpscenarios} would have problem with gauge invariance once coupled to the original $2$-group background fields. For continuous $2$-group symmetries a rigorous proof of \eqref{2grpscenarios} can be given using current algebra techniques \cite{Cordova:2018cvg}.

The symmetry breaking patterns appearing in \eqref{2grpscenarios} give rise to tight constraints on the allowed symmetries preserved by the vacuum. 
\begin{enumerate}
    \item[1.)] Namely, in $d>2$, it is impossible to realize a spontaneously broken 2-group with continuous $G\zero$ and finite $G\one$ in a conformal vacuum, since \emph{both} scenarios in \eqref{2grpscenarios} require the existence of Nambu--Goldstone bosons for the spontaneously broken $G\zero$ symmetry.
\end{enumerate}

Note that point $1.)$ above is unlike the discussion about CFTs and spontaneously broken finite $1$-form symmetries from section \ref{1form5dscft}. In principle, a conformal field theory might admit a spontaneously broken finite $1$-form symmetry since the existence of line operators obeying perimeter law is always compatible with conformal invariance.\footnote{A familiar example being four-dimensional $\cN=4$ $SU(N)$ super Yang--Mills theory.} However, our argument shows that there is a direct tension between conformal invariance in $d>2$ and a spontaneously broken $2$-group global symmetry whenever $G\zero$ is a continuous group and $G\one$ is finite.

We emphasize that a conformal field theory could have a decoupled $1$-form symmetry (see remarks in section \ref{1form5dscft}). In that event, a $2$-group can be unbroken, and the above does not apply.

It is also interesting to investigate the behavior of 2-groups under RG flows triggered by symmetry-preserving deformations. Consider an ultraviolet theory characterized by a 2-group global symmetry $\mathbb{G}_{UV} = (G_{UV}\zero, G_{UV}\one, 1, [\beta_{UV}])$ flowing to an infrared theory with $2$-group symmetry $\mathbb{G}_{IR} = (G_{IR}\zero, G_{IR}\one, 1, [\beta_{IR}])$. Even though $\mathbb G_{IR}$ and $\mathbb{G}_{UV}$ can be different, it is still possible to couple a theory to $\mathbb{G}_{UV}$ using backgrounds for $\mathbb{G}_{IR}$ \cite{Benini:2018reh}. A similar construction has been used already many times in this paper in the discussion of anomaly matching of sections \ref{subsec:3dSymmEnhancement} and \ref{subsec:5dAnomalyMatching}. This may only be achieved if there exist group homomorphisms $f_0, f_1$:
\be
f_0 : G_{UV}\zero \to G_{IR}\zero\ec \qquad f_1 : G_{UV}\one \to G_{IR}\one\ec
\ee
and if the following cohomological constraint holds:
\be\label{2grpconstraint}
f_1([\beta_{UV}]) = f^*_0([\beta_{IR}])\ec
\ee
where $f_0,f_1$ denote the maps on cohomology obtained from the group homomorphisms.
In a typical RG flow, $G_{UV}\one$ is included or equal to $G_{IR}\one$ i.e. $f_1$ is either a strict inclusion or an isomorphism. Note that this is not the case for $f_0$. More can be gleaned from the above constraint. For instance, if either $[\beta_{UV}]$ or $[\beta_{IR}]$ is non-trivial but the other one is not, then it follows that the maps $f_0,f_1$ cannot be isomorphisms. Said differently,
\begin{itemize}
    \item[2.)] If a UV theory with a non-trivial $2$-group global symmetry flows to an IR theory with trivial $2$-group, i.e. $[\beta_{IR}]=0$, then the IR theory must have emergent one-form symmetry such that $G_{UV}\one\subsetneq G_{IR}\one$. Conversely, if a UV theory with trivial 2-group, i.e. $[\beta_{UV}]=0$, flows to an IR theory with non-trivial $2$-group, then $G_{UV}\zero\neq G_{IR}\zero$.
\end{itemize}
The arguments from this section can be applied to some recent findings about the existence of $2$-group global symmetries in $5d$ SCFTs. The authors of \cite{Apruzzi:2021vcu} applied various geometrical tools of string theory to argue that the $5d$ $E_1$ SCFT admits a $2$-group of the form
\be\label{proposed2grp}
\mathbb{G}_{E_1} = (SO(3)_I\zero, \bZ_2\one,1,w_3)\ec
\ee
where the Postnikov class is identified with $w_3 = \textrm{Bock}(w_2(\cA_I))) \in H^3(BSO(3),\bZ_2)$ and $\textrm{Bock}$ is the Bockstein homomorphism associated to the short exact sequence $0 \to \Z_2 \to \Z_4 \to \Z_2 \to 0$. However, we believe that there are reasons to be cautious. 

The proposal \eqref{proposed2grp} is in tension with point $1.)$ above. The $\Z_2\one$ acts on the superconformal line defects of $\cT_{E_1}$. As such, it must be spontaneously broken. It follows that $\mathbb{G}_{E_1}$ \textit{too} should be spontaneously broken, but this is incompatible with conformal invariance as it would imply, see \eqref{2grpscenarios}, that $SO(3)\zero_I$ should be spontaneously broken as well.\footnote{As already remarked, to reach this conclusion we assume that $\Z_2\one$ acts on the superconformal line defects of $\cT_{E_1}$, see the comments in section \ref{subsec:5dAnomalyMatching}. However, it is possible that the geometric construction of $\cT_{E_1}$ gives rise to a decoupled $\Z_2\one$ symmetry.}

Another proposal of \cite{Apruzzi:2021vcu} is that $\cT_{E_1}$ has an ``extended" Coulomb branch where $SO(3)\zero_I$ is preserved and combines with $\bZ\one_2$ to form a $2$-group. This is compatible with both points $1.)$ and $2.)$. Indeed, on the Coulomb branch conformal invariance is spontaneously broken, and the low-energy effective theory is consistent with the breaking patterns \eqref{2grpscenarios}.\footnote{In fact, the low-energy theory is an IR free gauge theory, which in $d>4$ is a scale but not conformal invariant field theory.} Moreover, there is an additional Abelian symmetry generated by the (square of the) first Chern class of the dynamical $U(1)$ gauge field which is always available. So, $f_0$ is not an isomorphism.

Finally, observe that the supersymmetry-preserving deformation described by \eqref{flavormass} involves an explicit breaking of the global symmetry of the SCFT, so one cannot apply the argument of point $2.)$ above. Furthermore, there can be $5d$ RG flows exhibiting an emergent infrared $2$-group symmetry in the weakly-coupled gauge theory description. Some examples constructed adapting the constructions of \cite{Hsin:2020nts, Lee:2021crt} appear in \cite{Apruzzi:2021vcu}. Understanding whether this emergent symmetry structure imposes constraints on RG flows of $5d$ SCFTs is worth investigating.

\section{Supersymmetric 5d RG Flows}
\label{sec:SUSYRGFlows}

Supersymmetric gauge theories in five dimensions do not have an interesting IR behavior. Nonetheless, as we saw in section \ref{zeroform5dscft}, they can describe weakly-coupled phases obtained along RG flows from $5d$ SCFTs that admit flavor current multiplets.

Exploring the space of flavor mass deformation of a given $5d$ SCFT can in principle lead to many different infrared gauge theory descriptions. In this context, a question that was originally posed by the authors of \cite{Aharony:1997ju, Aharony:1997bh} is whether two different $5d$ supersymmetric gauge theory exhibiting isomorphic Coulomb branches could be characterized by the same UV fixed point (with identical set of local operators and OPE coefficients). From the infrared viewpoint it is not clear if this question is meaningful. However, the Coulomb phases of different-looking $5d$ gauge theories can arise in string theory from different resolutions of a unique singular limit, conjectured to correspond to a unique $5d$ SCFT.

The authors of \cite{Bergman:2013aca} were able to match the superconformal index (obtained by a localization computation of the Euclidean $S^4\times S^1$ supersymmetric partion function) of two different $5d$ supersymmetric gauge theories. This nontrivial computation has been dubbed (by an abuse of language) a ``UV duality,'' in which protected quantities of two different IR gauge theory match, see for instance \cite{Bergman:2014kza, Mitev:2014jza} for early examples. Furthermore, such analysis can also be generalized to include BPS defects, as shown in \cite{Gaiotto:2015una}, and to a number of gauge groups (for a recent review see \cite{Bhardwaj:2020gyu}).

Our goal in this section is to investigate if the recent developments on global generalized symmetries and their mixed 't Hooft anomalies, such as the ideas described in earlier sections, can be used to sharpen the proposed relations between different-looking IR gauge theories. More precisely:

\pagebreak

\begin{itemize}
    \item We consider examples of $5d$ SCFTs admitting a $G\zero_{UV}$ flavor current multiplet and (at least) two supersymmetry-preserving flavor mass deformations $\delta\scrL_A$ and $\delta\scrL_B$ leading to two distinct-looking IR gauge theories $\cT_A, \cT_B$, or to a unique gauge theory but with different identifications of the parameters.\footnote{It is currently not known how to classify all the inequivalent flavor mass deformations arising from a generic $5d$ SCFT with global symmetry.}
    \item Deforming by $\delta\scrL_{A,B}$ as in \eqref{flavormass} preserves a subgroup of the UV global symmetry $G\zero_{IR,A} = G\zero_{IR,B} \subset G\zero_{UV}$ along the entire RG flow that is visible in both IR gauge theories. We assume that a $1$-form symmetry $G\one_{A,B}$, if present at all, is never emergent in the RG flows we consider. This last assumption does not rely on the $1$-form symmetry being decoupled or not in the SCFT as we only require the existence of coupling to background fields.
    \item As emphasized in section \ref{subsec:5dAnomalyMatching}, the infrared supersymmetric gauge theories $\cT_A$ and $\cT_B$ can exhibit mixed 't Hooft anomalies between $G\zero_{IR,(A,B)}$ and $G\one_{IR,(A,B)}$. However, note that what happens to accidental $0$-form global symmetries is not relevant for this discussion.
\end{itemize}
The basic observation is that if $\cT_A$ and $\cT_B$ have different mixed 't Hooft anomalies, they cannot be obtained from a \textit{unique} $5d$ SCFT via the RG flow described above. This provides a stringent constraint that should refine the aforementioned conjectures regarding UV completions of different looking IR gauge theories. A similar comment has been made in \cite{Gukov:2020btk}.
Below, we will illustrate these general ideas using two related sets of concrete examples that highlight the most important features.

\subsection{\texorpdfstring{$SU(n+2)_{n+4}$}{SU(n+2)} and \texorpdfstring{$Sp(n+1)_{n\pi}$}{Sp(n+1)} Gauge Theories}

In this section we consider a $5d$ SCFT admitting a $U(1)_I\zero$ flavor current multiplet and two inequivalent $U(1)_I\zero$-preserving deformations \eqref{flavormass} parametrized by $m >0$ and $m<0$ where $m$ is the real flavor mass. Such theory should conjecturally arise at the origin of the Coulomb branch of two different-looking supersymmetric gauge theories described by (see \cite[(2.227)]{Bhardwaj:2020gyu})\footnote{In many available string theory constructions it is often hard to specify the topology of the gauge group. In what follows we take the simply connected Lie group corresponding to the Lie algebra determined geometrically.}
\beq
\label{eq:UVDualssusp}
\cT_{A}: SU(n+2)_{n+4} \longleftrightarrow \cT_{B}: Sp(n+1)_{n\pi} \, , \qquad m \geq 1 \, , 
\eeq
where the subscript indicates, respectively, the bare Chern--Simons level and the value of the discrete $\theta$ angle.

Both $\cT_A$ and $\cT_B$ have a conserved $U(1)\zero_I$ instantonic global symmetry. Moreover, using the results reviewed in Appendix \ref{app:OtherClassicalGroups}, we see that when $n$ is even the $1$-form symmetries $G\one_{A,B}$ are
\begin{equation}
\begin{split}
G\one_A &= \Z_{\gcd(n+2,n+4)}\one =
\Z_2\one\ec\\
G\one_B &= \Z_2\one\ec 
\end{split}
\end{equation}
while for odd $n$ there is no $1$-form symmetry for both theories.

We should now take into account all the possible 't Hooft anomalies involving global symmetries preserved by the flavor mass deformation. For $n=4k\ec k \in \N$, there is an anomaly on both sides (see Table \ref{Tab:MixedAnomalySummary}). On the left-hand side,  we find that the anomaly theory is
\beq\label{anomalyA}
A_6^{\cT_A}[\cA, \cB] = \exp \left( - 2\pi \ii \, \frac{1}{4} \int_{\cY_6}\frac{\rd\cA}{2\pi} \, \cP(\cB) \right)  \, ,
\eeq
and on the right-hand side, from \eqref{eq:SpNZTransformation}
\beq\label{anomalyB}
A_6^{\cT_B}[\cA', \cB'] = \exp \left( - 2\pi \ii \, \frac{1}{4}\int \frac{\rd \cA'}{2\pi} \, \cP(\cB') \right) \, ,
\eeq
so the two anomalies match upon matching the two symmetries. If $n=4k+2$, instead, the anomaly of $\cT_B$ remains non-trivial, but the anomaly of the left-hand side vanishes. This suggests that, if $G\one_A$ and $G\one_B$ are not emergent symmetries,  it is impossible to realize the proposal \eqref{eq:UVDualssusp} by deforming a unique ultraviolet fixed point. A solution to this issue is to consider the centerless groups with the same algebra as \eqref{eq:UVDualssusp}, rather than the simply connected versions.

One could also worry because $\cT_A$, being a gauge theory with a Chern--Simons term, could exhibit the additional cubic 't Hooft anomaly for $\Z_2\one$ found in \cite{Gukov:2020btk}, where $\cT_B$ would never have that. For pure $SU(N)_k$ it would take the form
\be\label{cubicthooft}
A_6^{\cT_A,\text{cubic}}[\cB] =  \exp \left( -2 \pi \ii \, \frac{k N (N-1)(N-2)}{6 \gcd(k,N)^3} \int \cB^3\right)\ec
\ee
where $\cB$ is a background field for $\Z_{\gcd(k,N)}\one$. However, for $N=n+2$, $k=n+4$, and even $n$, one can easily check that \eqref{cubicthooft} is trivial.

\subsection{Adding an Antisymmetric}

A related conjecture states that there exists a unique ultraviolet $5d$ SCFT  at the origin of the Coulomb branch of two different-looking gauge theories described by (see \cite[(2.229)]{Bhardwaj:2020gyu}):
\beq
\label{dualhol}
\cT_A: SU(n+2)_{\frac{n}{2}+5}+ \mathbf{\Lambda}^2 \longleftrightarrow \cT_B: Sp(n+1)_{n\pi} + \mathbf{\Lambda}^2 \, , \qquad n\geq 1 \, . 
\eeq
where $\mathbf{\Lambda}^2$ denotes a single hypermultiplet in the two-index antisymmetric representation of the gauge group. This example is particularly relevant, as the large $n$ limit of $\cT_B$ is amenable to a holographic description \cite{Brandhuber:1999np , Ferrara:1998gv}

As before, the UV $5d$ SCFT has (at least) a $U(1)\zero_I$ global symmetry and (at least) two inequivalent $U(1)_I\zero$ preserving deformations.\footnote{There is also a flavor symmetry associated to the hypermultiplet: for $SU(4)$ and the $Sp$ side, this is $Sp(1)$, whereas for $SU(n\geq 5)$ this is $U(1)$. The fact that the flavor symmetries on the two sides do not match means that we expect an enhancement of the quantum symmetry in the UV.} We follow the same conventions of the previous example and parametrize them by $m>0$ and $m<0$.

In order to determine the one-form symmetry of $\cT_A$ one has to work out the \emph{bare} Chern--Simons level and understand the effect of matter in antisymmetric representation. Similarly, for $\cT_B$, one should also consider the effect of the discrete $\theta$ angle. Following the procedure reviewed in appendix \ref{app:OtherClassicalGroups}, we find:
\be
G\one_A = G\one_B = \bZ\one_2\ec \quad n\in 2\bZ\ec
\ee
while $G\one_A$ and $G\one_B$ are both trivial if $m$ is odd. We can now discuss 't Hooft anomalies for both theories. It is important that on both sides the presence of matter in the antisymmetric representation does not break the $1$-form symmetry, so the discussion parallels that of the previous example. The theory $\cT_B$ exhibits a mixed 't Hooft anomaly $U(1)\zero_I \times \bZ\one_2$ anomaly of the same exact form of \eqref{anomalyB} which is always non-trivial. As before, $\cT_A$ has the mixed 't Hooft anomaly \eqref{anomalyA}, which is non-trivial if and only if $n\neq 4k+2$ with $k\in \N$, whereas the cubic one due to the Chern--Simons level again vanishes. It is clear that the anomalies of $\cT_A$ and $\cT_B$ do not match, exactly as in the previous example. We once more emphasize that, if $G\one_A$ and $G\one_B$ are not infrared emergent symmetries, it is impossible to realize the proposal \eqref{dualhol} by activating two inequivalent flavor current deformations of a unique UV fixed point.

Many gauge theories resulting from a relevant deformation of a $5d$ SCFT include matter content breaking completely the $1$-form symmetry. Nonetheless, it may still be that the flavor and instantonic symmetries participate in a mixed 't Hooft anomaly. This anomaly would again appear upon performing a large gauge transformation of the background gauge field for $U(1)\zero_I$, as in \eqref{eq:Variation5dPF}. However, the fractionalization of the instanton number would not be due the presence of a background gauge field for the center symmetry, but rather due to the presence of a background gauge field for the faithful flavor symmetry. This is the same phenomenon discussed in Section \ref{subsec:3dSymmEnhancement} in three dimensions, and has been discussed in \cite{BenettiGenolini:2020doj}. Indeed, one can use this anomaly as well in order to refine the proposed relations between infrared $5d$ gauge theories.

\bigskip
\begin{center}
\textbf{Acknowledgments}
\end{center}
\noindent We are grateful to Jeremías Aguilera-Damia, Fabio Apruzzi, Riccardo Argurio, Lakshya Bhardwaj, Francesco Benini, Matteo Bertolini, Mathew Bullimore, Stefano Cremonesi, Lorenzo Di Pietro, Eduardo García-Valdecasas, Zohar Komargodski, Francesco Mignosa, Kantaro Ohmori, Sakura Sch\"{a}fer-Nameki, David Tong and Yifan Wang for helpful discussions. We also thank the Galileo Galilei Institute for a stimulating environment during the workshop “Topological Properties of Gauge Theories and Their Applications to High-Energy and Condensed Matter Physics,” where part of this work was completed. The work of PBG has been supported by the Simons Foundation, by the STFC consolidated grant ST/T000694/1, and by the ERC Consolidator Grant N. 681908 “Quantum black holes: A macroscopic window into the microstructure of gravity.” LT has been partially supported by FNRS-Belgium (convention IISN 4.4503.15) and by funds from the Solvay Family. Any opinions, findings, and conclusions or recommendations expressed in this material are those of the authors and do not necessarily reflect the views of the funding agencies.

\newpage
\appendix

\section{Mixed Anomaly for Classical Gauge Groups}
\label{app:OtherClassicalGroups}

\subsection{Pure SYM}

The anomaly \eqref{an5d} was derived in Section \ref{subsec:5dAnomalyMatching} for the gauge group $SU(2)$. However, it is easy to generalise it to other classical gauge groups, as we here briefly summarize (see also \cite{BenettiGenolini:2020doj, Apruzzi:2021vcu}). 

Let $\mf{g}$ be a Lie algebra, and let $\tilde{G}$ be the unique simply connected Lie group with this Lie algebra. Its center is $Z(\tilde{G})$, and any group obtained as $G \equiv \tilde{G}/H$ with $H \subseteq Z(\tilde{G})$ has the same Lie algebra, and $\pi_1(\tilde{G}/H)\cong H$, whilst the center will be $Z(G) \cong Z(\tilde{G})/H$. Generically, there are gauge bundles of $G$ that may not be lifted to bundles of $\tilde{G}$: the obstruction is measured by a characteristic class $w \in H^2(BG,\pi_1(G)) \cong H^2(BG,H)$ (the ``discrete magnetic flux'' \cite{Witten:2000nv}). The same obstruction also measures the fractional part of the instanton number \cite{Cordova:2019bsd}\footnote{For the case of $\tilde{G}\cong Spin(4N)$ there are more subtleties, as we shall see.}
\beq
\label{eq:FractionalInstantonNumber}
\int_{\Sigma_4} \left( \cI - \frac{p_G}{q_G}\cP(w) \right) \in \Z \, , 
\eeq
where ${\Sigma_4}$ is a four-cycle in spacetime, the fractional number is
\beq
\label{eq:InstantonDensity}
\cI \equiv \frac{1}{4h^\vee}\Tr_{\rm adj} \frac{F}{2\pi} \wedge \frac{F}{2\pi} \, , 
\eeq
and $\cP$ is a short-hand notation for
\beq
\cP \equiv \begin{cases}
\text{Pontryagin square} \quad \cP &: H^2(-, \Z_{k}) \to H^4(-, \Z_{2k}) \ \text{$k$ even} \\
\text{cup product square} \quad \cup &: H^2(-, \Z_{k}) \to H^4(-, \Z_{k}) \ \text{$k$ odd} \\
\end{cases}
\eeq

Now consider pure SYM theory with gauge group $G$ with center $Z(\tilde{G})/H$: there is a one-form symmetry $(Z(\tilde{G})/H)^{(1)}$. We can turn on a background gauge field for this symmetry $\cB \in H^2(BG,H)$, thus introducing in the path integral bundles with group $G/(Z(\tilde{G})/H)$. Since these bundles have fractional instanton number \eqref{eq:FractionalInstantonNumber}, we find that, upon performing a large gauge transformation with unit winding number for the $U(1)_I^{(0)}$ symmetry, the partition function changes by
\beq
Z[\cA_I, \cB] \to Z[\cA_I, \cB] \exp \left( 2\pi \ii \frac{p_G}{q_G} \int_{\Sigma_4} \cP(\cB) \right) \, ,
\eeq
generalising \eqref{eq:Variation5dPF}. 

We shall now discuss the cases of the classical Lie groups to see whether the anomaly persists.

\subsection*{\texorpdfstring{$SU(N)$}{SU(N)}}

The center of $SU(N)$ is $\Z_N$, and we have subgroups $\Z_l$ for $l$ divisor of $N$. In this case
\beq
\begin{split}
w_2 \in H^2 \left( B\frac{SU(N)}{\Z_l} , \Z_l \right) \, , \qquad \frac{p_{SU(N)/\Z_l}}{q_{SU(N)/\Z_l}} = \frac{N(N-1)}{2l^2} \, .
\end{split}
\eeq
The transformation of the partition function is
\beq
\label{eq:SUNZTransformation}
Z[\cA, \cB] \to Z[\cA, \cB] \exp \left( 2\pi \ii \, \frac{N}{l}\frac{N-1}{2l} \int_{\Sigma_4}\cP(\cB) \right) \, .
\eeq
This may only be trivial if not only $l$, but also $l^2$ is a divisor of $N$.
For the simplest case $l=N$, we reduce to the case considered in \cite{BenettiGenolini:2020doj}. 

\subsection*{\texorpdfstring{$Sp(N)$}{Sp(N)}}

The center of $Sp(N)$ is $\Z_2$, and there are no non-trivial subgroups. In this case
\beq
\begin{split}
w_2 \in H^2 \left( B\frac{Sp(N)}{\Z_2} , \Z_2 \right) \, , \qquad \frac{p_{Sp(N)/\Z_2}}{q_{Sp(N)/\Z_2}} = \frac{N}{4} \, .
\end{split}
\eeq
The transformation of the partition function is
\beq
\label{eq:SpNZTransformation}
Z[\cA, \cB] \to Z[\cA, \cB] \exp \left( 2\pi \ii \, \frac{N}{4} \int_{\Sigma_4}\cP(\cB) \right) \, .
\eeq
This is trivial if $N$ is even, as already observed in \cite{BenettiGenolini:2020doj}.

\subsection*{\texorpdfstring{$Spin(2N+1)$}{Spin(2N+1)}}

The center of $Spin(2N+1)$ is $\Z_2$, and there are no non-trivial subgroups. In this case
\beq
\begin{split}
w_2 \in H^2 \left( B\frac{Spin(2N+1)}{\Z_2} , \Z_2 \right) \, , \qquad \frac{p_{Spin(2N+1)/\Z_2}}{q_{Spin(2N+1)/\Z_2}} = \frac{1}{2} \, .
\end{split}
\eeq
The transformation of the partition function is
\beq
Z[\cA, \cB] \to Z[\cA, \cB] \exp \left( 2\pi \ii \, \frac{1}{2} \int_{\Sigma_4}\cP(\cB) \right) \, ,
\eeq
which is always trivial.

\subsection*{\texorpdfstring{$Spin(4N+2)$}{Spin(4N+2)}}

The center of $Spin(4N+2)$ is $\Z_4$, so we may have one-form symmetry $\Z_2^{(1)}$ or $\Z_4^{(1)}$, which we summarize as $\Z_{2l}$ with $l=1, 2$. We have
\beq
w_2 \in H^2 \left( B \frac{Spin(4N+2)}{\Z_{2l}}, \Z_{2l} \right) \, , \qquad \frac{p_{Spin(4N+2)/\Z_{2l}}}{q_{Spin(4N+2)/\Z_{2l}}} = \frac{2N+1}{2l^2} \, .
\eeq
The transformation of the partition function is
\beq
Z[\cA, \cB] \to Z[\cA, \cB] \exp \left( 2\pi \ii \, \frac{2N+1}{2l^2} \int_{\Sigma_4}\cP(\cB) \right) \, .
\eeq
This is trivial for $l=1$, so for $Spin(4N+2)$ and $\Z_2^{(1)}$, whereas it's always non-trivial with $l=2$.

\subsection*{\texorpdfstring{$Spin(4N)$}{Spin(4N}}

The center of $Spin(4N)$ is $\Z_2^S\times \Z_2^C$, where the two factors have different actions on the spinor representation. We can take four different quotients
\beq
\label{eq:SpinQuotients}
\begin{aligned}
\frac{Spin(4N)}{\Z^S_2} \equiv Ss(4N) \, , &\qquad
&\frac{Spin(4N)}{\Z_2^C} \equiv Sc(4) \, , \\ 
\frac{Spin(4N)}{\Z^V_2} \cong SO(4N) \, , &\qquad &\frac{Spin(4N)}{(\Z_2^S \times \Z_2^C)} \cong \frac{SO(4N)}{\Z_2} \, ,
\end{aligned}
\eeq
where $\Z_2^V$ is the diagonal subgroup of $\Z_2^S\times \Z_2^C$.
The background gauge fields for the two $\Z_2$ factors are denoted by $\cB^S, \cB^C$. For the first two quotients, leading to the semispin theories, we have
\beq
\frac{p_{Ss(4N)}}{q_{Ss(4N)}} = \frac{p_{Sc(4N)}}{q_{Sc(4N)}} = \frac{N}{4} \, ,
\eeq
so the anomaly is the same as that of $Sp(N)$ case, and it is trivial for even $N$
\beq
Z[\cA, \cB^{S/C}] \to Z[\cA, \cB^{S/C}] \exp \left( 2\pi \ii  \, \frac{N}{4} \int_{\Sigma_4}\cP(\cB^{S/C}) \right) \, .
\eeq
For the $SO(4N)$ theory, we should take $\cB^S = \cB^C \equiv \cB$ and
\beq
\frac{p_{SO(4N)}}{q_{SO(4N)}} = \frac{1}{2} \, ,
\eeq
so the anomaly is trivial, which leads us to conclude that for all $SO(n)$ theories there is no anomaly, even though the center of $SO(2n)$ is $\Z_2$, so there would still be a left-over one-form symmetry. Finally, for the $SO(4N)/\Z_2$ theory, the relation \eqref{eq:FractionalInstantonNumber} is modified to
\beq
\int \left( \cI - \frac{N}{4}\cP(\cB^S+\cB^C) - \frac{1}{2} \cB^S \cup \cB^C \right) \ \in \Z \, ,
\eeq
so
\beq
\begin{split}
Z[\cA, \cB^S, \cB^C] &\to Z[\cA, \cB^S, \cB^C] \exp \left[ 2\pi \ii \, \int_{\Sigma_4} \left( \frac{N}{4}\cP(\cB^S+\cB^C) - \frac{1}{2} \cB^S \cup \cB^C \right) \right] \\
&= Z[\cA, \cB^S, \cB^C] \exp \left[ 2\pi \ii \, \frac{N}{4} \int_{\Sigma_4} \cP(\cB^S+\cB^C) \right] \, ,
\end{split}
\eeq
and it vanishes for even $N$.

\medskip

A summary of the possible mixed 't Hooft anomalies between the center and instantonic symmetries is in Table \ref{Tab:MixedAnomalySummary}.

\begin{table}[tb]
\begin{center}
\begin{tabular}{ccc}
\toprule
Group & $1$-form & Anomaly? \\
\midrule
\rowcolor{lightgray} $SU(N)$ & $\Z_l$ & Yes, if $l^2\nmid N$ \\
\rowcolor{lightgray} $Sp(2N+1)$ & $\Z_2$ & Yes \\
$Sp(2N)$ & $\Z_2$ & No \\
$Spin(2N+1)$ & $\Z_2$ & No \\
$Spin(4N+2)$ & $\Z_2$ & No \\
\rowcolor{lightgray} $Spin(4N+2)$ & $\Z_4$ & Yes \\
\rowcolor{lightgray} $Spin(8N+4)$ & $\Z_2^{S/C}$ & Yes \\
$Spin(8N)$ & $\Z_2^{S/C}$ & No \\
$Spin(4N)$ & $\Z_2^V$ & No \\
\rowcolor{lightgray} $Spin(8N+4)$ & $\Z_2^S\times \Z_2^C$ & Yes \\
$Spin(8N)$ & $\Z_2^S \times \Z_2^C$ & No \\
\bottomrule
\end{tabular}
\end{center}
\caption{Summary of the mixed 't Hooft anomalies $U(1)\zero_I\times Z(G)\one$.}
\label{Tab:MixedAnomalySummary}
\end{table}

\subsection{Chern--Simons Terms and Matter}

There are additional effects that may reduce the one-form symmetry of the theory, such as the presence of a Chern--Simons term and matter (see also \cite{Morrison:2020ool, Albertini:2020mdx}).

In $SU(N)$ SYM with $N\geq 3$, it is possible to construct a five-dimensional Chern--Simons term\footnote{Of course, it is also possible to construct a Chern--Simons term for the Abelian gauge theory, but here we focus on the non-abelian case.}
\be
\label{eq:cs5}
\SL_{\rm CS} = -\frac{\ii k}{24 \pi^2}\Tr\left(A \wedge F \wedge F + \frac{\ii}{2} A \wedge A \wedge A \wedge F - \frac{1}{10} A \wedge A \wedge A \wedge A \wedge A\right)\ed
\ee
The \textit{bare level} $k$ must be an integer in order to preserve gauge invariance. The presence of a Chern--Simons term at bare level $k$ explicitly breaks the one-form symmetry from $\Z_N\one$ to $\Z\one_{\gcd(k,N)}$. There are various ways of showing this, including adjoint Higgsing \cite{BenettiGenolini:2020doj} or, for theories constructed from singular compactifications in M-theory, geometric considerations \cite{Morrison:2020ool, Albertini:2020mdx} (see also the analysis in \cite{Gukov:2020btk}). This implies that one may only turn on background gauge fields for the $\Z\one_{\gcd(k,N)}$ subgroup, and therefore, may only include in the path integrals bundles in $SU(N)/\Z_{\gcd(k,N)}$.

The presence of matter affects both the Chern--Simons term, and the 1-form symmetry. Recall that quantizing a five-dimensional fermion leads to an anomaly between gauge symmetry and parity, which can also be measured in the three-point function of currents \cite{Alvarez-Gaume:1984zst}. Integrating out the fermion with a large mass shifts the Chern--Simons level in the Lagrangian, and the integration with masses with different sign leads to theories where the Chern--Simons level differs by $1$. We should then choose a convention for this effect, and we follow \cite{Closset:2018bjz, Morrison:2020ool} in choosing the $U(1)_{-\frac{1}{2}}$ quantization. That is, if we start with $SU(N)_k$ SYM with a hypermultiplet in the fundamental, giving a large positive mass to the spinor in the hyper does not change the level in the infrared theory, whereas a large negative mass leads to $SU(N)_{k-1}$ SYM. For a fermion in a generic representation $\mathbf{R}$, a large positive mass does not shift the Chern--Simons level in the infrared, whereas a large negative mass shifts the Chern--Simons level by the cubic Casimir of $\mathbf{R}$, $A_{\mathbf{R}}$. 

Following standard conventions, we label a $SU(N)_{k}$ SYM theory with the \textit{effective level} $k$. Different from the bare Chern--Simons level, the effective level may also be rational, and we write it as
\beq
k \equiv k_{\textrm{bare}} - \frac{1}{2}\sum_{i=1}^m A_{\mathbf{R}_i} \, , 
\eeq
where $k_{\textrm{bare}}$ is the bare Chern--Simons level, and $i$ runs over the fermions in representation $\mathbf{R}_i$. For instance, the theory
\beq
SU(N)_{k} + N_F \mathbf{F} + N_{\Lambda^2} \mathbf{\Lambda}^2 \, ,
\eeq
where $\mathbf{\Lambda}^2$ denotes the two-index antisymmetric representation, has bare Chern--Simons level
\beq
\label{eq:BareChernSimonsLevelMatterSYM}
k_{\textrm{bare}} = k + \frac{N_F}{2} + N_{\Lambda^2} \frac{N-4}{2} \, .
\eeq

Matter may also directly break the one-form symmetry of the theory. Namely, it breaks the one-form center symmetry to the maximal subgroup of $Z(G)\one$ under which matter is uncharged. Concretely, for $Z(G)\cong \Z_n$, a representation of $\Z_n$ is determined by the image of $1$, which must be a root of unity. Thus, each of the $n$ roots of unity leads to an inequivalent representation $\rho_j$ such that $\rho_j(1)= \e^{2\pi \ii j/n}$, and the charge is $j$, which is thus defined modulo $n$. An extended table of charges of some representations can be found in \cite{Bhardwaj:2020phs}. For our purposes, we shall need the following charges under the center symmetry
\begin{center}
\begin{tabular}{ccc}
\toprule
Gauge group & Representation & Charge \\
\midrule
$SU(N)$ & $\mb{F}$ & $1 \mod N$ \\
\vspace{0.2cm}
 & $\mb{\Lambda}^2$ & $2 \mod N$ \\
$Sp(N)$ & $\mb{F}$ & $1 \mod 2$ \\
\vspace{0.2cm}
 & $\mb{\Lambda}^2$ & $ 0 \mod 2$ \\
\bottomrule
\end{tabular}
\end{center}
Notice that hypermultiplets in the fundamental representation break completely the $1$-form center symmetry, whereas the adjoint representation always has trivial charge.

Finally, it is possible to also include a discrete $\theta$ angle, which is valued in $\pi_4(G)$ and weighs non-trivial gauge configurations in the path integral \cite{Douglas:1996xp}. The only simple simply connected Lie group with non-trivial $\pi_4$ is $Sp(n)$, for which $\pi_4(Sp(n))\cong \Z_2$. Thus, for $Sp(n)$ SYM, we may also have a $\Z_2$-valued $\theta$ angle: a non-trivial $\theta$ angle explicitly breaks completely the center one-form. Similar to the convention for the Chern--Simons level, integrating out a hypermultiplet in the fundamental of $Sp(n)_\theta$ with positive mass does not change $\theta$, whereas integrating out with negative large mass changes $\theta$ by $\pi$.

\newpage
%
\renewcommand\refname{\bfseries\large\centering References\\ \vspace{-0.4cm}
\addcontentsline{toc}{section}{References}}
\bibliographystyle{utphys.bst}
{
\bibliography{Anomaly5D.bib}%
}
	
\end{document}